%% file: GaussianFAR.tex
\documentclass[aps,prd,twocolumn,preprintnumbers,nofootinbib,superscriptaddress,amsmath]{revtex4-1}
\usepackage{graphicx}
\usepackage{amsmath}
\usepackage{amssymb}
\usepackage{amsfonts}
\usepackage{hyperref}
\usepackage{url}
\usepackage{color}
\usepackage{bm}
\usepackage{array}
\usepackage{placeins}

\usepackage[table]{xcolor}
\definecolor{lightgray}{gray}{0.9} % table alternating line colors
\def\mystrut{\vrule height 9.3pt depth 3.1pt width 0pt}

\newcommand{\ssout}[1]{}

\def\pastro{p_{\rm astro}}
\def\rNLO{\rho_{\rm NLO}}

\newenvironment{Far_table}{\setlength{\tabcolsep}{0pt}}{}

\begin{document}
\makeatother

\preprint{IFT-UAM/CSIC-22-105}

\title{The False Alarms induced by Gaussian Noise in Gravitational Wave Detectors}

\author{Gonzalo Morr\'as}
\affiliation{Instituto de F\'isica Te\'orica UAM/CSIC, Universidad Aut\'onoma de Madrid, Cantoblanco 28049 Madrid, Spain}
\author{Jose Francisco Nu\~no Siles}
\affiliation{Instituto de F\'isica Te\'orica UAM/CSIC, Universidad Aut\'onoma de Madrid, Cantoblanco 28049 Madrid, Spain}
\author{Juan Garc\'ia-Bellido}
\affiliation{Instituto de F\'isica Te\'orica UAM/CSIC, Universidad Aut\'onoma de Madrid, Cantoblanco 28049 Madrid, Spain}
\author{Ester Ruiz Morales}
\affiliation{Departamento de F\'isica Aplicada, ETSIDI, Universidad Polit\'ecnica de Madrid, 28012 Madrid, Spain}
\affiliation{Instituto de F\'isica Te\'orica UAM/CSIC, Universidad Aut\'onoma de Madrid, Cantoblanco 28049 Madrid, Spain}

\date{\today}

\begin{abstract}
\noindent 
Gaussian noise is an irreducible component of the background in gravitational wave (GW) detectors. Although stationary Gaussian noise is uncorrelated in frequencies, we show that there is an important correlation in time when looking at the matched filter signal to noise ratio (SNR) of a template, with a typical autocorrelation time that depends on the template and the shape of the noise power spectral density (PSD). Taking this correlation into account, we compute from first principles the false alarm rate (FAR) of a template in Gaussian noise, defined as the number of occurrences per unit time that the template's matched filter SNR goes over a threshold $\rho$. We find that the Gaussian FAR can be well approximated by the usual expression for uncorrelated noise, if we replace the sampling rate by an effective sampling rate that depends on the parameters of the template, the noise PSD and the threshold $\rho$. This results in a minimum SNR threshold that has to be demanded to a given GW trigger, if we want to keep events generated from Gaussian noise below a certain FAR. 
We extend the formalism to multiple detectors and to the analysis of GW events. We apply our method to the GW candidates added in the GWTC-3 catalog, and discuss the possibility that GW200308\_173609 and GW200322\_091133 could be generated by Gaussian noise fluctuations.
\end{abstract}
\maketitle

\section{Introduction}
\label{sec:intro}

A century after their theoretical derivation from General Relativity~\cite{Einstein1916}, Gravitational Waves (GWs) are now routinely detected by the laser interferometers of the LIGO-Virgo-KAGRA collaboration~\cite{AdvLIGO_design,AdvVirgo_design,Kagra_design}. Their amplitude is so small that their detection above instrumental and environmental noise requires sophisticated pipelines~\cite{SPIIR,PyCBC_description_2015,PyCBC,GstLAL,MBTA,cWB}, which look for signals in the data with various methods. 
These pipelines have to be designed to reject noise from very common non-Gaussian transient sources of noise (also known as glitches)~\cite{DetChar_O2O3}, while being computationally efficient to search for events in a wide range of parameters within an affordable amount of time. 

In the case of modeled searches for GWs from Compact Binary Coalescences (CBCs), templates from a predefined template bank are compared with the data at all times to find where a GW signal can be present. The likelihood that the observed data contains a GW signal is quantified by computing a pipeline-specific ranking statistic, defined in such a way that the larger its value the more it favors the signal hypothesis versus the noise hypothesis. If the detector noise were purely Gaussian, it can be proved that the optimal ranking statistic for a signal of known form would be the matched filter SNR~\cite{Helstrom_TheoryOfSignalDetection}.  
However, the search pipelines that actually look for GWs use ranking statistics that, although based on the SNR, introduce corrections to consider the presence of non-Gaussian glitches which can give sizeable spurious SNR values. The corrections are usually based on signal consistency tests, a common example being the use of $\chi^2$~\cite{PyCBC_Ranking} to weigh down the SNR.

In order to assign a significance to the candidate events in terms of their ranking statistic, the pipelines need to find the background distribution of the ranking statistic for the bank of templates. This is estimated in a data driven way, usually by running the search on the time-shifted strain of the different interferometers, so that coincidences become not physical and the triggers obtained this way represent an estimate of the background noise. The false alarm rate (FAR) of an event is then defined by the search pipeline as the rate of background triggers over the whole bank of templates with ranking statistic equal to or higher than the one observed for the event.
Therefore, the FAR can give us an idea of how likely it is for noise to generate an event. Intuitively, for a total observation time $T_\text{obs}$, any trigger that has $\text{FAR} \geq 1/T_\text{obs}$ is compatible with being generated by noise, while $\text{FAR} \ll 1/T_\text{obs}$ disfavors the noise hypothesis.

In searches for GWs, the FAR estimates can differ several orders of magnitude among different pipelines~\cite{GWTC-3}, given that the FAR usually has an exponential dependence on the ranking statistic. Therefore, small variations in how the data is processed, what templates are used or what is looked for to rank the events in the different pipelines, can result in orders of magnitude discrepancies in the estimation of the FAR.

Moreover, the FAR does not contain any information about the foreground. To take this into account, together with the astrophysical prior knowledge, the $\pastro$, was introduced~\cite{pastroOG}. The rationale behind $\pastro$ is to give the Bayesian probability that a candidate is from astrophysical origin under a model for the foreground rates $f(x,\vec{\theta})$ and background rates $b(x,\vec{\theta})$ that depend on the ranking statistic $x$ and the template parameters $\vec{\theta}$.
A threshold value of $\pastro>0.5$ was required for any candidate event to be included in the GWTC-3 catalog~\cite{GWTC-3}. The estimated expected contamination from events of terrestrial origin is $\sim$ 10–15$\%$, or $\sim$ 4–6 events. In the same fashion as the FAR, the $\pastro$ for a given event can be very different between pipelines and presents large uncertainties, especially around $\pastro \sim 0.5$~\cite{MBTA_pastro_uncertinties}. 

As a consequence of the application of this threshold to enter the GWTC-3 Catalog, some events were accepted with FAR values greater than $1/T_\text{obs} \sim 2 \mathrm{yr}^{-1}$. One example is GW200322\_091133~\cite{GWTC-3} with FAR $>400 \mathrm{yr}^{-1}$, which, upon further investigation with Bayesian Parameter Estimation (PE), was found to have low SNR ($\leq 8.5$) and multimodal posterior distributions of its parameters. Since the likelihood used in PE is approximately proportional to $\exp(\mathrm{SNR}^2/2)$, in events with small SNR the likelihood will not have a large enough peak so as to dominate the posterior, and there will be prior-dominated modes.

All these difficulties may prompt one to think that these candidate events with low SNR values might come from noise fluctuations. The noise and GW signal hypotheses are usually compared locally using the \textit{Bayes factor} \cite{GW_BayesFactor}. However, this number says nothing about how often we expect noise to generate a signal as ``loud'' as the observed one. This has motivated us to question whether we could aim to obtain a theoretical lower bound on the false alarm rate of an event, independently of all the complexities involved in the search pipelines. We start from the idea that Gaussian noise is always an irreducible component of the background in GW detectors~\cite{Abbott:2016xvh, GuideToLIGOVirgoNoise}, and generates a rate of false alarms that could be calculated analytically. In the case in which non-Gaussianities are also present in the strain, more false alarms will be induced~\cite{Detchar_for_GW150914}, as matches will occur more easily for a given template, thus making our estimate assuming only Gaussian noise a lower bound on their FAR, and thus an upper bound on their significance. 

In this paper, we propose a new method to derive a local statistical measure of the significance of an event. The main idea will be to give a theoretical estimate of how often we would expect Gaussian noise colored with the local PSD to produce a fluctuation that matches a specific template with the same or higher SNR than the one observed.
In Sec.~\ref{sec:SNR_FAR} we develop the framework to compute the FAR for a given template in Gaussian noise from a single detector and study its dependence on different parameters for CBC templates. In Sec.~\ref{sec:SNR_FAR:ManyDet} we extend the formalism to compute the FAR of a template when multiple detectors are online. In Section~\ref{sec:SNR_FAR:GW_events} we  show  how to apply our statistical method to fluctuations observed in the strain and in Sec.~\ref{sec:SNR_FAR:GW_events:LVK} we use it on the O3b events included in GWTC-3. Finally in section~\ref{sec:Conclusions} we present our conclusions.

\section{The false alarm rate of a template in a single detector}
\label{sec:SNR_FAR}

In this section we want to determine, given a template $h(t)$, how much time of stationary Gaussian noise $n(t)$, from a given detector, we would have to look at, on average, to obtain a match with a signal to noise ratio (SNR) greater than some threshold $\rho$.

\begin{figure*}[t!]
\begin{center}
\includegraphics[width=\textwidth]{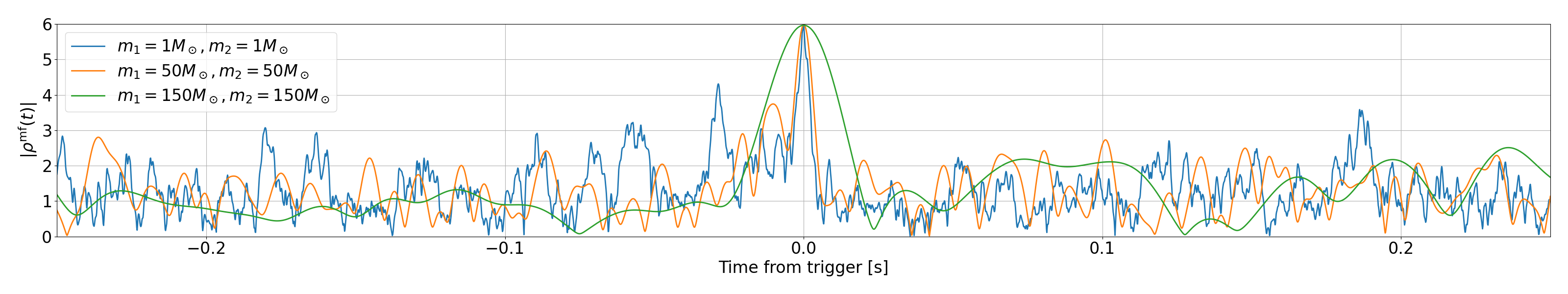}
\end{center} 
\caption{Simulation of the modulus of the matched filter SNR $|\rho^\mathrm{mf}(t)|$ for Gaussian noise generated using Advanced LIGO design sensitivity \cite{AdvLIGO_design} and \texttt{IMRPhenomPv2} \cite{Khan_2019} templates of masses $m_1 = m_2 = 1M_\odot$, $m_1 = m_2 = 50 M_\odot$ and $m_1 = m_2 = 150 M_\odot$. Each template is matched with different noise realizations until we obtain a trigger of $|\rho^\mathrm{mf}(t_\mathrm{trig})| \sim 6$. We plot 0.5s around this trigger.}
\label{fig:Gaussian_snr_t_realization}
\end{figure*}

In general the noise will have zero mean, $\langle \tilde{n} \rangle = 0$, and assuming that it is stationary, the different Fourier modes are uncorrelated, 
\begin{equation}
    \langle \tilde{n}^{*}(f) \tilde{n}(f') \rangle \equiv \frac{1}{2} S_{n}(f) \delta(f-f') \,,
    \label{eq:PSD_def}
\end{equation}
which can be seen as the definition of the noise power spectral density (PSD) $S_{n}(f)$. If we assume that the noise is Gaussian, it is characterized completely by the fact that it has zero mean and a variance given in Eq.~\eqref{eq:PSD_def}. Using the PSD we can define the following inner product,
\begin{equation}
    \langle a, b \rangle = 4 \int_{f_{\mathrm{min}}}^{f_{\mathrm{max}}} \frac{\tilde{a}^{*}(f)\tilde{b}(f)}{S_{n}(f)} df \,,
    \label{eq:inner_product}
\end{equation}
where tildes denote Fourier transform. This inner product can be used to write down the usual definitions \cite{IntroBayesInference_GWs} of the optimal SNR:
\begin{equation}
    \rho^\mathrm{opt} = \sqrt{\langle h , h \rangle} \,, 
    \label{eq:SNR_opt_def}
\end{equation}
and the matched filter SNR:
\begin{equation}
    \rho^\mathrm{mf} = \frac{\langle h , s \rangle}{\rho^\mathrm{opt}} \,,
    \label{eq:SNR_mf_def}
\end{equation}
where $s(t)$ is the detector output strain, which in our case we will assume to be given by stationary Gaussian noise $n(t)$ with PSD $S_{n}$. Under this assumption, it can be proved that $\rho^\mathrm{mf}$ is a complex normal random variable (i.e. a Gaussian with unit dispersion, $\sigma=1$) \cite{Maggiore_Vol1}:
\begin{equation}
    p(\rho^\mathrm{mf}) d \mathrm{Re} \rho^\mathrm{mf} d \mathrm{Im} \rho^\mathrm{mf} = \frac{1}{2 \pi} e^{ -\frac{1}{2}|\rho^\mathrm{mf}|^2 } d \mathrm{Re} \rho^\mathrm{mf} d \mathrm{Im} \rho^\mathrm{mf} \, .
    \label{eq:SNR_mf_gausian_distribution}
\end{equation}
\noindent and the real part of the matched filter SNR is the optimum quantity to rank the significance of events for a signal of known form under the assumption of Gaussian noise \cite{Helstrom_TheoryOfSignalDetection}. This quantity is very closely related to the likelihood ratio for the signal vs Gaussian noise hypotheses, which is the Bayes factor for a signal of known intrinsic parameters. However, it is common to be in the situation in which the global phase of the GW can be changed arbitrarily and does not contain any astrophysical information~\cite{LALInference_PE}. This is the case in a quasicircular compact binary coalescence, when we ignore higher order modes and precession. Even when including them, the global phase can typically be neglected since it is highly degenerate with other parameters such as polarization, location in the sky and the azimuthal angle separating the spin vectors of the component BHs. In these cases we will want to ignore the global phase of the GW in the search by taking as our ranking statistic the absolute value of the matched filter SNR:
\begin{equation}
    |\rho^\mathrm{mf}| = \sqrt{\mathrm{Re}(\rho^\mathrm{mf})^2+\mathrm{Im}(\rho^\mathrm{mf})^2}\,,
    \label{eq:ranking_statistic}
\end{equation}
which is invariant under global phase transformations $\tilde{h}(f) \to \tilde{h}(f) e^{i \phi_g}$. Indeed, the SNR usually used in searches is $|\rho^\mathrm{mf}|$ \cite{PyCBC_Ranking_2020} since it is equivalent to automatically finding the global phase $\phi_g$ of the GW that maximizes $\mathrm{Re}(\rho^\mathrm{mf})$. Because of this we will choose $|\rho^\mathrm{mf}|$ as our ranking statistic in this paper. Defining $\tilde{h}(f)$ as the Fourier transform  of the template $h(t)$, we can use the following property:
\begin{equation}
    \mathcal{F}(h(t')) = 
    \tilde{h}(f) e^{-2 \pi i f (t'-t)} \,,
    \label{eq:h_at_other_time}
\end{equation}
and compute the matched filter signal to noise ratio, Eq.\eqref{eq:SNR_mf_def}, at all times as
\begin{equation}
    \rho^\mathrm{mf}(t) = \frac{4}{\rho^\mathrm{opt}} \int_{f_\mathrm{min}}^{f_\mathrm{max}} \!\!\! df\, \frac{\tilde{h}^{*}(f) \tilde{n}(f)}{S_n (f)} e^{2 \pi i f t} \, .
    \label{eq:SNR_mf_t}
\end{equation}
\noindent where we assume that the strain only contains Gaussian noise. At any fixed point in time, $\rho^\mathrm{mf}(t)$ of Eq.~\eqref{eq:SNR_mf_t} will behave as a complex normal variable from Eq.~\eqref{eq:SNR_mf_gausian_distribution} and the probability of obtaining a value of $|\rho^\mathrm{mf}|$ greater than $\rho$ will be:
\begin{align}
    P(|\rho^\mathrm{mf}|>\rho) & = \frac{1}{2\pi} \int_0^{2\pi} \!\!\!
    d \,\mathrm{arg}(\rho^\mathrm{mf}) \int_{\rho}^\infty \!\! |\rho^\mathrm{mf}| d |\rho^\mathrm{mf}| e^{-\frac{1}{2} |\rho^\mathrm{mf}|^2}  \nonumber \\
    & = e^{-\frac{1}{2} \rho^2} \, .
    \label{eq:P_SNR_mf_gtr_thr}
\end{align}
A naive computation to estimate the rate of false alarms with $|\rho^\mathrm{mf}|>\rho$ would be to multiply this probability by the number of trials per unit time, which in the case that different times were independent, would just be the sampling rate of the detector:
\begin{align}
    \mathrm{FAR}_\mathrm{naive} = \frac{1}{\Delta t_\mathrm{samp}} e^{-\rho^2/2} \, .
    \label{eq:FAR_naive}
\end{align}
However, this would be  incorrect because the value of $|\rho^\mathrm{mf}(t)|$ at different times is correlated. 
The problem can be explicitly seen in Fig.~\ref{fig:Gaussian_snr_t_realization}, where we have generated Gaussian noise from Advanced LIGO at design sensitivity \cite{AdvLIGO_design} and 
computed $|\rho^\mathrm{mf}(t)|$ using Eq.~\eqref{eq:SNR_mf_t} with \texttt{IMRPhenomPv2} \cite{Khan_2019} templates of the specified masses. Each template is matched with different noise realizations until we obtain a trigger of $|\rho^\mathrm{mf}(t_\mathrm{trig})| \sim 6$, which we show in Fig.~\ref{fig:Gaussian_snr_t_realization}. The correlation between different times manifests itself in the fact that $|\rho^\mathrm{mf}(t)|$ is a smooth function, where the smoothing time scale will be related to the autocorrelation time, and we observe that it depends on the template mass. In particular, the larger the mass, the larger the autocorrelation time will be. This correlation of $|\rho^\mathrm{mf}(t)|$ at different times has a direct effect on the False Alarm Rate (FAR), defined as the average time between peaks with $|\rho^\mathrm{mf}|>\rho$, since the smoother the function $|\rho^\mathrm{mf}(t)|$ is, the less peaks per second it will have, thus reducing the rate of false alarms. Assuming that the sampling rate of the detector is sufficiently fine to see $|\rho^\mathrm{mf}(t)|$ as a smooth function, we will demonstrate in the rest of this section that the effect of the correlations will be to replace the sampling rate of the detector $1/\Delta t_\mathrm{samp}$ in Eq.~\eqref{eq:FAR_naive} by an effective sampling rate that depends on the template, the noise PSD and the threshold $\rho$.

\subsection{Probabilistic derivation of the FAR}

The autocorrelation of $\rho^\mathrm{mf}(t)$ can be quantified by computing the covariance between the values of $\rho^\mathrm{mf}(t)$ at different times, assuming that the strain only contains Gaussian noise:
\begin{align}
    &\Gamma(t,t') = \frac{1}{2}\langle \rho^\mathrm{mf}(t) \rho^\mathrm{mf}(t')^{*}\rangle =  \nonumber \\
    & = 8 \left\langle  \! \int_{f_\mathrm{min}}^{f_\mathrm{max}} \hspace{-5mm} df \int_{f_\mathrm{min}}^{f_\mathrm{max}} \hspace{-5mm} df' \frac{\tilde{h}^{*}(f) \tilde{h}(f') \tilde{n}^{*}(f') \tilde{n}(f)}{S_n (f) S_n(f')
    \ (\rho^\mathrm{opt})^2} 
    e^{2\pi i(f t - f' t')} \right\rangle \nonumber \\
    & = 8 \int_{f_\mathrm{min}}^{f_\mathrm{max}} \hspace{-5mm} df \int_{f_\mathrm{min}}^{f_\mathrm{max}} \hspace{-5mm} df' \frac{\tilde{h}^{*}(f) \tilde{h}(f') \langle \tilde{n}^{*}(f')\tilde{n}(f) \rangle}
    {S_n (f) S_n(f')\ (\rho^\mathrm{opt})^2} 
    e^{2\pi i(f t - f' t')} \nonumber \\
    & = \Gamma(t-t') = \frac{4}{(\rho^\mathrm{opt})^2} \int_{f_\mathrm{min}}^{f_\mathrm{max}} \! \! \! d f  \frac{|\tilde{h}(f)|^2}{S_n(f)} e^{2\pi i f (t-t')} \,, 
    \label{eq:SNR_mf_t_autocorr}
\end{align}
where we have used Eq.~\eqref{eq:PSD_def} and that $\langle \rho^\mathrm{mf}(t) \rangle = 0$. 
We observe in Eq.~\eqref{eq:SNR_mf_t_autocorr} that for $t=t'$ we have $\Gamma(0)=1$, as expected from the fact that $\rho^\mathrm{mf}(t)$ is a complex normal variable at any specific point in time. 
In general
$\Gamma(t-t')$ will be non-negligible for $t \neq t'$, 
so the value of the SNR at two different times will be correlated. %Therefore, if we study 
If we consider the SNR at two different points separated by a time $\Delta t$, and define $\rho^\mathrm{mf}(t) \equiv \rho_1^c$ and $\rho^\mathrm{mf}(t+\Delta t) \equiv \rho_2^c$, from Eq.~\eqref{eq:SNR_mf_t_autocorr} we have that their joint probability distribution will be given by the following bivariate complex Gaussian:
\begin{equation}
    p(\rho_1^c, \rho_2^c) = \frac{\exp \left\{- \frac{|\rho_1^c|^2 + |\rho_2^c|^2 - 2 \mathrm{Re}(\Gamma(\Delta t) \rho_1^{c *} \rho_2^c)}{2(1-|\Gamma(\Delta t)|^2)} \right\}}{(2 \pi)^2 (1-|\Gamma(\Delta t)|^2)} \,,
    \label{eq:bivariate_cgaussian}
\end{equation}
Using this expression we can compute the two-point false alarm probability (FAP$_2$), that is, the probability that either $\rho_1$ or $\rho_2$ are greater than some SNR threshold $\rho$,
\begin{equation}
    \mathrm{FAP}_2 = P(\rho_1 > \rho \, \cup \, \rho_2 > \rho) \,.
    \label{eq:FAP2_0}    
\end{equation}

An in depth study of this quantity is made in appendix~\ref{sec:anex:FAP_bcG}, where we find expressions to compute it numerically and to analytically approximate it to arbitrary order. To understand how FAP$_2$ behaves, and to gain intuition on how the FAP of more variables will behave, it is interesting to discuss its limiting behaviours. When the separation between the two points is large ($\Delta t \to \infty$), the correlation between them vanishes ($|\Gamma(\Delta t)| \to 0$) meaning that FAP$_2$ becomes the FAP of two uncorrelated variables, that is, $\mathrm{FAP}_2(|\Gamma(\Delta t)|=0) = 2e^{-\rho^2/2} - e^{-\rho^2}$. As the points get closer together ($\Delta t \to 0$) the correlation increases ($|\Gamma(\Delta t)| \to 1$), and FAP$_2$ will decrease due to correlation effects until the correlation is maximal ($|\Gamma(\Delta t)| = 1$), when the two variables will behave as a single one and $\mathrm{FAP}_2(|\Gamma(\Delta t)|=1) = e^{-\rho^2/2}$.

In the real setup of a GW experiment, we are interested in determining the false alarm probability for N points separated by a sampling time $\Delta t$ each. If we define ${\rho_k \equiv |\rho^\mathrm{mf}(t + k \Delta t)|}$, this FAP is given by:
\begin{align}
    \mathrm{FAP} & = P\left( \bigcup_{n=1}^N \rho_n > \rho\right) = 1 - P\left( \bigcap_{n=1}^N \rho_n < \rho \right) \nonumber \\
     & = 1 - P(\rho_1 < \rho) \prod_{k=2}^{N} P\Big( \rho_{k} < \rho \ \Big| \bigcap_{n=1}^{k-1}\rho_n<\rho \Big)\, ,
   \label{eq:FAPN}
\end{align}
\noindent where $P(A|B)$ denotes the conditional probability of $A$ given $B$  and in the last equality we have used the multiplication rule of probability. To compute Eq.~\eqref{eq:FAPN} we will thus need ${P(\rho_2 < \rho | \rho_1 < \rho)}$. This can be computed in terms of the FAP$_2$ defined in Eq.~\eqref{eq:FAP2_0}:
\begin{align}
    P(\rho_2 < \rho | \rho_1 < \rho) &=  \frac{P(\rho_1 < \rho \cap \rho_2 < \rho)}{P(\rho_1 < \rho)} \nonumber \\
    & = \frac{1 - \mathrm{FAP}_2(\rho, \Delta t)}{1 - e^{-\rho^2/2}} \nonumber \\ 
    & \approx 1 - (\mathrm{FAP}_2(\rho, \Delta t) - e^{-\rho^2/2}) \, ,
    \label{eq:Conditional_FAP2}
\end{align}
where in the last equality we have assumed that ${e^{-\rho^2/2} \ll 1 }$ (which is true for $\rho \gtrsim 3$). In order to compute Eq.~\eqref{eq:FAPN} we also need to calculate $P(\rho_k < \rho \,|\,\rho_1 < \rho \, \cap \rho_2 < \rho \cap ... \cap \rho_{k-1}<\rho)$. We can determine this conditional probability in an approximate way by assuming that it depends only on the nearest neighbor, that is:
\begin{align}
    P & (\rho_k < \rho \,|\, \rho_1 < \rho \cap  \rho_2 < \rho \cap ... \cap \rho_{k-1} < \rho)  \nonumber \\
    & \approx P(\rho_k < \rho \,|\, \rho_{k-1} < \rho ) = P(\rho_2 < \rho \,|\, \rho_1 < \rho)\, ,
    \label{eq:Neirest_Neighbor_approx}
\end{align}
\noindent where in the last equality we have just used the translation invariance of the problem. The Nearest Neighbor approximation of Eq.~\eqref{eq:Neirest_Neighbor_approx} will only be valid in the case in which the sampling time $\Delta t$ is large enough such that second neighbor effects can be neglected, which could be taken into account by replacing the approximation of Eq.~\eqref{eq:Neirest_Neighbor_approx} by $P(\rho_3 < \rho | \rho_2 < \rho \cap \rho_1 < \rho)$.

Introducing Eqs.~\eqref{eq:Conditional_FAP2}, \eqref{eq:Neirest_Neighbor_approx} into Eq.~\eqref{eq:FAPN} and assuming that $\mathrm{FAP}_2 - e^{-\rho^2/2} \ll 1$, we have:
\begin{align}
        \mathrm{FAP} & \approx 1 - (1-e^{-\rho^2/2})\left[ 1 - \left(\mathrm{FAP}_2(\rho, \Delta t) - e^{-\rho^2/2}\right) \right]^{N-1} \nonumber \\
        & \approx 1 - \exp\left\{ - N \left[\mathrm{FAP}_2(\rho, \Delta t) - e^{-\rho^2/2}\right] \right\} \nonumber \\
        & \approx 1 - \exp\left\{ - \frac{T_\mathrm{obs}}{\Delta t} \left[\mathrm{FAP}_2(\rho, \Delta t) - e^{-\rho^2/2}\right] \right\} \, ,
    \label{eq:FAPN_approx}
\end{align}
\noindent where $T_\mathrm{obs}$ is the observing time on which we are computing the FAP, which we assume to be long enough so that $N =  T_\mathrm{obs}/\Delta t \gg 1$. 

To obtain a quantity that is independent of the observing time, we define the false alarm rate (FAR), which is the average number of false alarms per unit time. As we see in Fig.~\ref{fig:Gaussian_snr_t_realization}, the autocorrelation of the SNR has the effect of clustering its values in peaks. Though each peak of $|\rho^\mathrm{mf}(t)|$ has many sample times over the threshold, which naively could count as false alarms, it is important to realize that each peak should be counted as a single false alarm, that is, we have to find the number of {\sl uncorrelated} false alarms which are thus Poisson distributed. This is an important point, given that if each sample time that is over the SNR threshold $\rho$ were counted as a false alarm, we would obtain the naive FAR of Eq.~\eqref{eq:FAR_naive}, since looking at individual points the probability is given by Eq.~\eqref{eq:P_SNR_mf_gtr_thr}, and we would greatly overestimate the FAR. 

By the definition of the FAR, the mean of the Poisson distribution describing the number of uncorrelated false alarms will be $\lambda = T_\mathrm{obs} \mathrm{FAR}$, assuming an observing time $T_\mathrm{obs}$. Therefore, the probability of having $k$ false alarms is:
\begin{align}
    p(k)  = \frac{(T_\mathrm{obs} \mathrm{FAR})^k}{k!}e^{-T_\mathrm{obs} \mathrm{FAR}} \, .
    \label{eq:Poisson_dist_uncorr_false_alarms}    
\end{align}
Since the FAP is the probability of having one or more false alarms, it is given by:
\begin{align}
    \mathrm{FAP}  & =\sum_{k=1}^\infty p(k) = 1 -p(0) = 1 - \exp\{-T_\mathrm{obs} \mathrm{FAR}\} \, .
    \label{eq:FAP_of_FAR}    
\end{align}
By comparing Eq.~\eqref{eq:FAPN_approx} and Eq.~\eqref{eq:FAP_of_FAR}, we immediately deduce the following relation between the FAR and the FAP:
\begin{align}
    \mathrm{FAR}_2(\rho, \Delta t)  =  \frac{1}{\Delta t} \left[\mathrm{FAP}_2(\rho, \Delta t) - e^{-\rho^2/2}\right] \, ,
    \label{eq:FAR2_exact}    
\end{align}
\noindent where we add the subscript 2 to highlight that this FAR has been computed taking into account only nearest neighbors. 

\subsection{Evaluation of the FAR of a template}

In order to further elaborate the expression of the FAR for a given template in
Eq.~\eqref{eq:FAR2_exact}, we need to study the $\mathrm{FAP}_2(\rho, \Delta t)$ more in depth. In the case in which the detector has a high enough sampling rate, we can assume that $|\rho^\mathrm{mf}(t)|$ is a continuous function, as is the case in Fig.~\ref{fig:Gaussian_snr_t_realization}. This will be a very good approximation in LIGO-Virgo, where the data is taken at a sampling rate of $1/\Delta t_\mathrm{samp} = 16384$~Hz. In this case, instead of interpreting $\Delta t$ as the sampling time of the detector, we leave it as a free parameter, as we imagine that the function $|\rho^\mathrm{mf}(t)|$ can be resampled arbitrarily. We will want to make $\Delta t \to 0$, to obtain the result for when $|\rho^\mathrm{mf}(t)|$ is continuously sampled, but if $\Delta t$ is too small, the nearest neighbor approximation will stop being valid. The effect of the farther neighbors will be to reduce the number of effective trials. This compensates the increase in the number of sampling points in such a way that the exact FAR with all correlations taken into account will be smaller than the FAR from the nearest neighbor approximation, that is:
\begin{align}
    \mathrm{FAR}(\rho, \Delta t_\mathrm{samp}) & \leq \mathrm{FAR}_2(\rho, \Delta t_\mathrm{samp}) \, .
    \label{eq:FARexact_upper bound_FAR2}
\end{align}

With this in mind, we approximate the FAR of Eq.~\eqref{eq:FAR2_exact} for $\Delta t \to 0$, which from Eq.~\eqref{eq:SNR_mf_t_autocorr} is equivalent to $|\Gamma(\Delta t)| \to 1$. We can do this by introducing in Eq.~\eqref{eq:FAR2_exact} the expression for FAP$_2$ of Eq.~\eqref{eq:FAP2_approx_LO} found in Appendix.~\ref{sec:anex:FAP_bcG}, keeping only next to leading order terms in $1-|\Gamma(\Delta t)|$ and assuming that $\rho^2 \gg 1$:
\begin{align}
    \mathrm{FAR}_2 \approx  \frac{e^{-\rho^2/2}}{\Delta t} \mathrm{Erf}\left[\frac{\rho\sqrt{1-|\Gamma(\Delta t)|}}{2}\right] \,.
    \label{eq:FAR_approx_alpha}    
\end{align}
Since we are interested in the limit $\Delta t \to 0$, we can substitute $\Gamma(\Delta t)$ by its Taylor expansion around $\Delta t = 0$, which using the definition in Eq.~\eqref{eq:SNR_mf_t_autocorr} will be given by:
\begin{align}
    \Gamma(\Delta t) & = \frac{4}{(\rho^\mathrm{opt})^2} \int_{f_\mathrm{min}}^{f_\mathrm{max}} \! \! \! d f  \frac{|\tilde{h}(f)|^2}{S_n (f)} e^{2\pi i f \Delta t} \nonumber \\
    & = \frac{4}{(\rho^\mathrm{opt})^2} \int_{f_\mathrm{min}}^{f_\mathrm{max}} \! \! \! d f  \frac{|\tilde{h}(f)|^2}{S_n (f)} \sum_{k=0}^\infty \frac{(2\pi i f \Delta t)^k}{k!} \nonumber \\ 
    & = \sum_{k=0}^{\infty} i^k \frac{C_k}{k!} (\Delta t)^k  \, ,
    \label{eq:alpha_taylor}    
\end{align}
where $C_k$ are real constants defined as
\begin{equation}
    C_k = \frac{4}{(\rho^\mathrm{opt})^2} \int_{f_\mathrm{min}}^{f_\mathrm{max}} \hspace{-1mm} df \, (2 \pi f)^k  \frac{|\tilde{h}(f)|^2}{S_n (f)}   \, .
    \label{eq:C_k}
\end{equation}
To leading order in $\Delta t$, we then have that $|\Gamma(\Delta t)|$ will be given by:
\begin{align}
    |\Gamma(\Delta t)| = 1 &- \frac{1}{2} \left( C_2 - C_1^2 \right) (\Delta t)^2 \, ,
    \label{eq:abs_alpha_taylor}    
\end{align}
\noindent where we have used that $C_0 = 1$. Substituting the expansion for $|\Gamma(\Delta t)|$ of Eq.~\eqref{eq:abs_alpha_taylor} into Eq.~\eqref{eq:FAR_approx_alpha} and keeping terms in $\Delta t$ up to leading order, we obtain:
\begin{equation}
    \mathrm{FAR}_2(\rho, \Delta t) \approx \frac{e^{-\rho^2/2}}{\Delta t}\mathrm{Erf}\left[ \frac{\sqrt{\pi}}{2} \rho C \Delta t  \right] \,,
    \label{eq:FAR2_approx_LO}
\end{equation}
where for simplicity we have defined: 
\begin{equation}
        C  \equiv \sqrt{\frac{C_2 - C_1^2}{2 \pi}}\, ,
        \label{eq:C_coef}
\end{equation}
\noindent which is always a real quantity, since $C_2 - C_1^2 \geq 0$.~\footnote{We can explicitly prove that $C_2 - C_1^2 \geq 0$ and gain some intuition on $C$, if we realize that
\begin{equation}
    g(f) = 
    \begin{cases}
       \frac{4}{(\rho^\mathrm{opt})^2} \frac{|\tilde{h}(f)|^2}{S_n (f)} &\; f_\mathrm{min} < f < f_\mathrm{max}\\
       0 &\; \mathrm{else} \\
    \end{cases}
    \label{eq:eff_prob_dist}
\end{equation}
\noindent can be interpreted as a probability distribution function, since it is always non-negative and it is normalized (i.e. $\int_{-\infty}^\infty g(f) df =1$). Using this probability distribution function, we then observe that $C$ is simply given by:
\begin{align}
        C  & = \sqrt{2 \pi (\mathbb{E}_g[f^2] - \mathbb{E}_g[f]^2)} = \sqrt{2 \pi \mathbb{E}_g\left[(f  - \mathbb{E}_g[f])^2 \right]} = \sqrt{2 \pi} \,\sigma_f \, .
        \label{eq:C_coef_from_pdf}    
\end{align}
\noindent where $\mathbb{E}_g[X]$ denotes the expectation value of $X$ in $g$, $\sigma_f$ is the standard deviation of the frequency $f$ in $g$, and from the second equality we explicitly see that the argument of the square root is always positive. From Eq.~\eqref{eq:C_coef_from_pdf} we then observe that $C$ will be directly related with the bandwidth, that is, how spread out in frequencies is $g(f)$. Therefore, the more broadband our detector and signals are, the larger $C$ will be in general.}
This is a necessary condition given by the fact that $C_2 - C_1^2$ is the leading order coefficient in the Taylor expansion of $|\Gamma(\Delta t)|$ (see Eq.~\eqref{eq:abs_alpha_taylor}) and we know that $|\Gamma(\Delta t)| \leq 1$.

From Eq.~\eqref{eq:FAR2_approx_LO} we have that in the limit $\Delta t \to 0$:
\begin{equation}
    \mathrm{FAR}_2(\rho, 0) = C\,\rho\,e^{-\rho^2/2} \, .
    \label{eq:FAR2_0}
\end{equation}
The way to interpret the result of Eq.~\eqref{eq:FAR2_0} is that even if we consider the separation between points to tend to 0, the FAR will not diverge, as we would have naively deduced from Eq.~\eqref{eq:FAR_naive}. The correlation between the neighboring points will regularize the FAR to the finite value of Eq.~\eqref{eq:FAR2_0}. 

\begin{figure}[t!]
\begin{center}\hspace*{-2mm}
\includegraphics[width=0.495\textwidth]{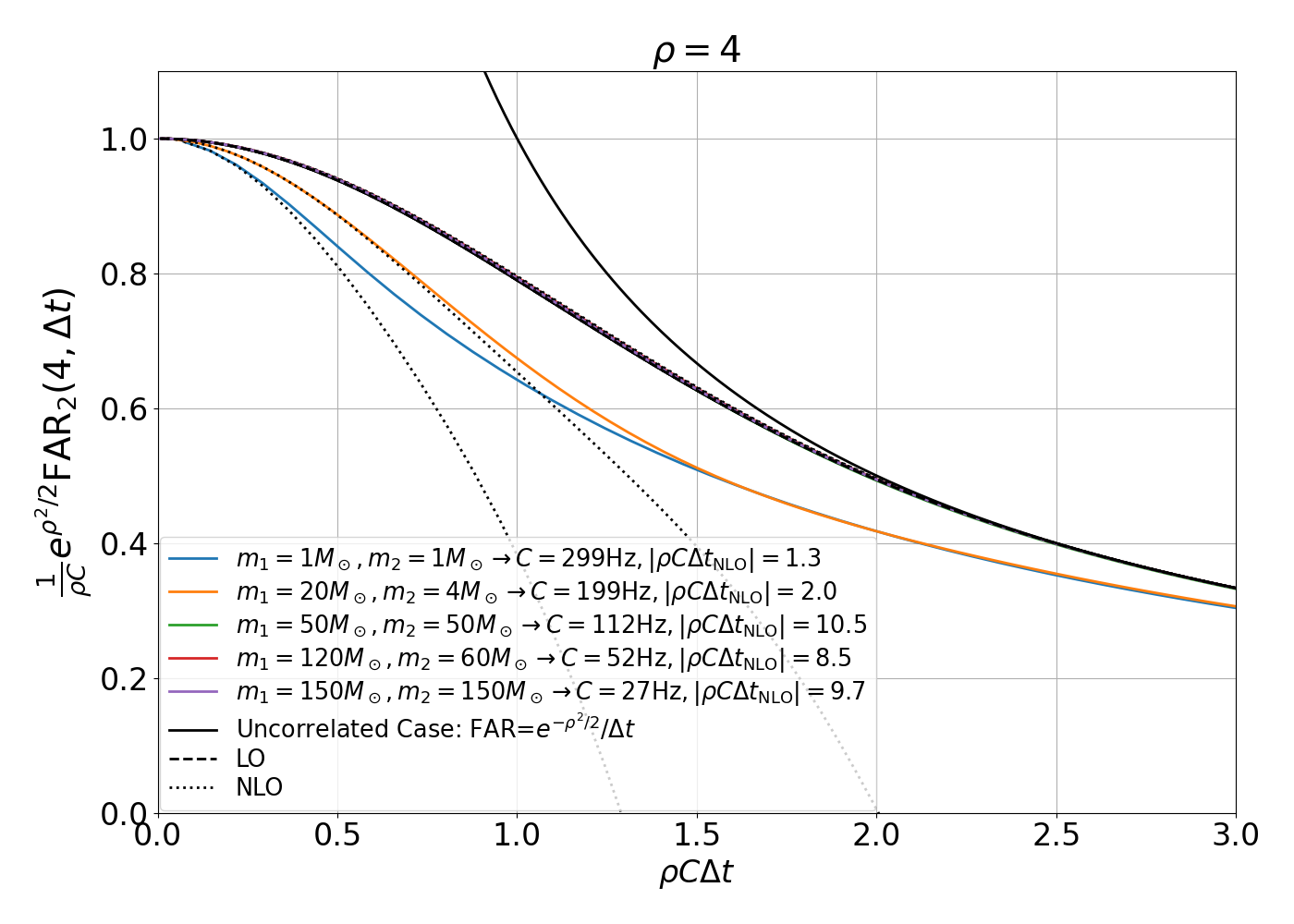}
\end{center} 
\caption{The FAR$_2$ for \texttt{IMRPhenomPv2} \cite{Khan_2019} templates of different masses, assuming Advanced LIGO at design sensitivity \cite{AdvLIGO_design} and setting the threshold SNR $\rho=4$. The FAR is computed using the exact expression (Eq.~\eqref{eq:FAR2_exact}), the leading-order expression of Eq.~\eqref{eq:FAR2_approx_LO} and the NLO expression of Eq.~\eqref{eq:FAR2_approx_NLO}, where the integrals in frequency are always computed between $f_\mathrm{min} = 20$~Hz and $f_\mathrm{max} = 2048$~Hz to mimic normal GW analysis. We normalize the FAR to its value at 0 separation and the time to make the LO approximation of cases appear the same. The uncorrelated case of Eq.~\eqref{eq:FAR_naive} is also plotted.}
\label{fig:FAR2_autocorr}
\end{figure}

This can be seen in Fig.~\ref{fig:FAR2_autocorr}, where we show the FAR$_2$ for \texttt{IMRPhenomPv2} \cite{Khan_2019} templates of different masses, assuming Advanced LIGO at design sensitivity \cite{AdvLIGO_design}. The FAR is computed using the exact expression (Eq.~\eqref{eq:FAR2_exact}), the leading order (LO) expression of Eq.~\eqref{eq:FAR2_approx_LO} and the next-to-leading order (NLO) expression of Eq.~\eqref{eq:FAR2_approx_NLO}, which will be discussed in the next subsection. For the cases of large masses ($m_{1,2} = 50 M_\odot$, $m_{1,2} = 150 M_\odot$ and $m_1 = 120 M_\odot, m_2=60 M_\odot$), we have that the leading order expression gives an accurate representation of the exact result, as can be seen from the fact that the lines for the three high mass cases are on top of each other and on top of their corresponding LO and NLO approximations. This is no longer true for the low mass cases of $m_{1,2} = 1 M_\odot$ and $m_1 = 20 M_\odot, m_2=4 M_\odot$, where the FAR decreases faster than expected at high values of $\Delta t$ due to correlation tails at this high $\Delta t$. To describe this deviation from the LO result, we will have to take into account higher order corrections in $\Delta t$, which will be discussed in the next subsection.

The fewer trials we do, the smaller the FAR should be. Therefore the FAR is a monotonously decreasing function of $\Delta t$, and $\mathrm{FAR}_2(\rho, \Delta t_\mathrm{samp}) \leq \mathrm{FAR}_2(\rho, 0)$, which can correctly be seen in Fig.~\ref{fig:FAR2_autocorr}. Using this together with Eq.~\eqref{eq:FARexact_upper bound_FAR2} we obtain
\begin{align}
    \mathrm{FAR}(\rho, \Delta t_\mathrm{samp}) \leq \mathrm{FAR}_2(\rho,  0)  = \rho C e^{-\rho^2/2} \, .
    \label{eq:FARexact_upper bound}
\end{align}

We expect that the result of Eq.~\eqref{eq:FARexact_upper bound} will be a very tight upper bound, and thus a good approximation of the exact FAR in the case that the NLO corrections are small, since these are related with the length of the correlations and thus the importance of the next-to-near neighbors. 

To study the validity of this result we will simulate the problem at hand. In particular, we will simulate the FAP by generating many chunks of simulated Gaussian noise from Advanced LIGO at design sensitivity \cite{AdvLIGO_design} of duration $T_\mathrm{obs}=512$~s. We directly compute the probability to have a trigger with $|\rho^\mathrm{mf}|>\rho$ by performing matched filtering on the noise using a GW template and dividing the number of chunks where we find a match with $|\rho^\mathrm{mf}|>\rho$ by the total number of chunks analyzed. From this FAP we can obtain the FAR simply by inverting Eq.~\eqref{eq:FAP_of_FAR}:
\begin{equation}
    \mathrm{FAR} = \frac{1}{T_\mathrm{obs}} \log\left( \frac{1}{1 - \mathrm{FAP}} \right) \, .
    \label{eq:FAR_of_FAP}
\end{equation}

\begin{figure}[t!]
\begin{center}\hspace*{-2mm}
\includegraphics[width=0.5\textwidth]{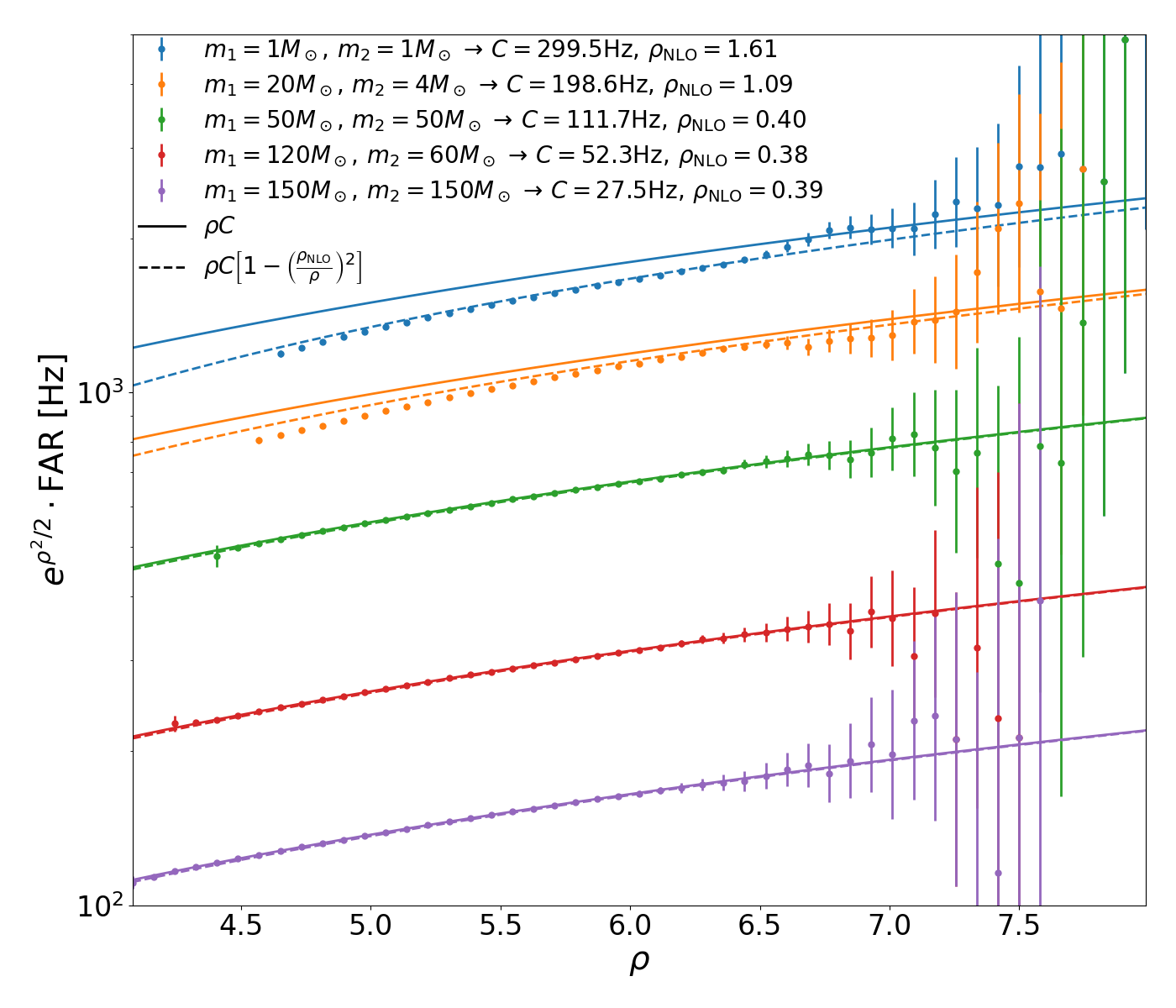}
\end{center} 
\caption{Comparison between the simulated and predicted FAR for five different \texttt{IMRPhenomPv2} templates. The simulation is done using 15 million chunks of 512 s of simulated Gaussian noise generated from Advanced LIGO at design sensitivity \cite{AdvLIGO_design}. We directly compute the probability to have a trigger with $|\rho^\mathrm{mf}|>\rho$ by performing matched filtering between $f_\mathrm{min} = 20$~Hz and $f_\mathrm{max} = 2048$~Hz with the corresponding GW template and dividing the number of chunks where we find a match with $|\rho^\mathrm{mf}|>\rho$ by the total number of chunks analyzed. The error on the FAP is computed using the Wilson score 90\% confidence interval \cite{WilsonScoreInterval}. Introducing this FAP in Eq.~\eqref{eq:FAR_of_FAP} (using $T_\mathrm{obs}=$512 s) we obtain the FAR plotted with dots, whose error bars represent the 90\% confidence interval. For the theory curves, the corresponding values of $C$ are computed with Eq.~\eqref{eq:C_coef}) and Eq.~\eqref{eq:C_k}.
}
\label{fig:fsamp_eff}
\end{figure}

In Fig.~\ref{fig:fsamp_eff} we show the FAR computed in this way from the simulation of the FAP and multiplied by $e^{\rho^2/2}$ to extract the exponential decay behavior and make visualization easier. The matched filter is done with five different \texttt{IMRPhenomPv2} templates with the same masses as the ones used in Fig.~\ref{fig:FAR2_autocorr}. We have observed that indeed, Eq.~\eqref{eq:FARexact_upper bound} is always satisfied and $\mathrm{FAR}_2(\rho, 0)$ is an upper bound of $\mathrm{FAR}_N(\rho, \Delta t_\mathrm{samp})$ within the error. As was discussed previously, this is a tight upper bound in the case in which the NLO corrections are small, deviating by less that 1 part in 1000 for the larger masses ($m_{1,2} = 50 M_\odot$, $m_{1,2} = 150 M_\odot$ and $m_1 = 120 M_\odot, m_2=60 M_\odot$). In the cases where the NLO corrections are important ($m_{1,2} = 1 M_\odot$ and $m_1 = 20 M_\odot, m_2=4 M_\odot$) we can observe that even though Eq.~\eqref{eq:FARexact_upper bound} is still a good upper bound, it is not so tight any more. Nonetheless, the maximum relative error between the upper bound and the exact value always stays below 15\% and decreases towards larger values of the SNR threshold $\rho$. We thus confirm that a good approximation of the FAR is:
\begin{align}
    \mathrm{FAR} = C\,\rho\, e^{-\rho^2/2} \, .
    \label{eq:FARtot_approx}
\end{align}

Comparing this expression with the value of the naive FAR that we derived at the beginning in Eq.~\eqref{eq:FAR_naive}, we have that, as anticipated, the sampling time of the experiment is naturally replaced by an effective sampling time for which we can obtain the same result as for uncorrelated points. This effective sampling rate depends on the threshold $\rho$ and on the template and noise PSD via the coefficient $C$:
\begin{equation}
    \Delta t_\mathrm{eff} = \frac{1}{\rho C}
    \label{eq:dt_dec}
\end{equation}

Consistently computing corrections to this result, we would have to take into account the effect of next-to-leading order corrections. We do this in the next subsection.

\subsection{NLO corrections to the FAR of a template}

We will start by studying the next-to-leading-order (NLO) corrections to the expression for FAR$_2$ found in Eq.~\eqref{eq:FAR2_approx_LO}. For this we now substitute in Eq.~\eqref{eq:FAR2_exact} the expression for FAP$_2$ of Eq.~\eqref{eq:FAP2_approx_NLO} found in Appendix.~\ref{sec:anex:FAP_bcG}, keeping NLO terms in $1-|\Gamma(\Delta t)|$ and assuming that $\rho^2 \gg 1$:
\begin{align}
    \mathrm{FAR}_2 \approx  \frac{e^{-\rho^2/2}}{\Delta t} \mathrm{Erf}\bigg\{&\frac{\rho\sqrt{1-|\Gamma(\Delta t)|}}{2} \bigg(1 + \frac{1-|\Gamma(\Delta t)|}{4} \bigg)\bigg\} \, .
    \label{eq:FAR_approx_alpha_NLO}    
\end{align}
And when considering the Taylor expansion of $|\Gamma(\Delta t)|$ we now keep up to quartic terms, that is:
\begin{align}
    |\Gamma(\Delta t)| = 1 &- \frac{1}{2} \left( C_2 - C_1^2 \right) (\Delta t)^2 \nonumber \\ 
    & + \frac{1}{24}(C_4 - 4 C_1 C_3 + 6 C_1^2 C_2 - 3 C_1^4) (\Delta t)^4 \, ,
    \label{eq:abs_alpha_taylor_p}    
\end{align}
Introducing this Taylor expansion into Eq.~\eqref{eq:FAR_approx_alpha_NLO} and keeping up to leading order terms, we have:
\begin{equation}
    \mathrm{FAR}_2(\rho, \Delta t) \! \approx \! \frac{e^{-\rho^2/2}}{\Delta t}\mathrm{Erf}\!\left\{\! \frac{\sqrt{\pi}}{2} \rho C \Delta t \left( 1 - \frac{(\Delta t)^2}{(\Delta t_\mathrm{NLO})^2} \right) \! \right\} \, ,
    \label{eq:FAR2_approx_NLO}
\end{equation}
\noindent where we have introduced $\Delta t_\mathrm{NLO}$ as the characteristic time for which when $\Delta t \ll |\Delta t_\mathrm{NLO}|$ we can neglect higher order effects. In terms of $C_k$, it will be given by:
\begin{equation}
    (\Delta t_\mathrm{NLO})^2 = \frac{24 (C_2 - C_1^2)}{C_4 - 4 C_1 C_3 - 3 C_2^2 + 12 C_1^2 C_2 - 6 C_1^4} \, .
    \label{eq:Delta_t_NLO}
\end{equation}
Looking again at Fig.~\ref{fig:FAR2_autocorr} where the NLO FAR$_2$ of Eq.~\eqref{eq:FAR2_approx_NLO} is compared in  with the LO expression (Eq.~\eqref{eq:FAR2_approx_LO}) and with the exact expression (Eq.~\eqref{eq:FAR2_exact}), we can observe that the NLO corrections are not important for the high mass systems ($m_{1,2} = 50 M_\odot$, $m_{1,2} = 150 M_\odot$ and $m_1 = 120 M_\odot, m_2=60 M_\odot$), since $|\rho C \Delta t_\mathrm{NLO}| \gg 1$. However, for the low mass cases of $m_{1,2} = 1 M_\odot$ and $m_1 = 20 M_\odot, m_2=4 M_\odot$, which have $|\rho C \Delta t_\mathrm{NLO}| \sim O(1)$, we can see that the higher order corrections in $\Delta t$ are important. In these cases, the tails of the correlation are relatively longer, and so the FAR decreases faster than expected as a function of $\Delta t$, which is accurately described by the NLO corrections as long as $\Delta t \lesssim \Delta t_\mathrm{NLO}$.

We also want to obtain a more accurate formula for the Gaussian FAR than the one in Eq.~\eqref{eq:FARtot_approx}. To consistently compute corrections to the result of Eq.~\eqref{eq:FARtot_approx}, we would have to take into account the effect of farther neighbors in Eq.~\eqref{eq:Neirest_Neighbor_approx}. Nonetheless, doing this becomes very complicated rather quickly. Instead, a heuristic way to take into account the next to leading order corrections can be found by imposing that these preserve the same behavior as the leading order term of Eq.~\eqref{eq:FAR2_approx_LO}, which we have seen gives a very good description when higher orders can be neglected. We can imagine that at $\Delta t_\mathrm{eff}/2$ there will be a sampling point whose correlation we are neglecting when we resample $|\rho^\mathrm{mf}(t)|$. We will then impose that the correlation $|\Gamma(\Delta t)|$ at this point has the same value as in the case where we only consider the leading order term in the Taylor expansion of Eq.~\eqref{eq:abs_alpha_taylor}:
\begin{equation}
    \left|\Gamma\left(\frac{\Delta t_\mathrm{eff}}{2}\right)\right| = 1 -  \frac{\pi}{4 \rho^2} \, .
    \label{eq:Delta_t_dec_NLO_heuristic_equation_0}
\end{equation}
Using the next to leading order expansion for $|\Gamma(\Delta t)|$ on the left hand side, we obtain:
\begin{equation}
    1 - \frac{\pi}{4} \left(C \Delta t_\mathrm{eff}\right)^2  + \frac{\pi}{2} \rNLO^2 \left(C \Delta t_\mathrm{eff}\right)^4 = 1 -  \frac{\pi}{4 \rho^2} \, .
    \label{eq:Delta_t_dec_NLO_heuristic_equation_1}
\end{equation}
where for convenience we have defined $\rNLO$ in the following way
\begin{equation}
    \rNLO = \sqrt{\frac{\pi ( C_4 - 4 C_1 C_3 + 6 C_1^2 C_2 - 3 C_1^4)}{48 (C_2-C_1^2)^2}} \, .
    \label{eq:rho_NLO}
\end{equation}
Solving Eq.~\eqref{eq:Delta_t_dec_NLO_heuristic_equation_1} for $\Delta t_\mathrm{eff}$, keeping only leading-order terms in $\rNLO/\rho$, we obtain:
\begin{equation}
    \frac{1}{\Delta t_\mathrm{eff}^\mathrm{NLO}} = \rho C \left[1 - \left(\frac{\rNLO}{\rho} \right)^2 \right] \, .
    \label{eq:Delta_t_dec_NLO_heuristic}
\end{equation}

This heuristic result is compared in Fig.~\ref{fig:fsamp_eff} with the simulated value. Although we have to keep in mind that it has not been derived in a consistent way, we can observe that it closely follows the behavior of the deviations from Eq.~\eqref{eq:FARtot_approx} for the cases of $m_{1,2} = 1 M_\odot$ and $m_1 = 20 M_\odot, m_2=4 M_\odot$ for which the corrections are important. Eq.~\eqref{eq:Delta_t_dec_NLO_heuristic} will thus be a useful model to understand how these deviations behave. As expected, the heuristic corrections of Eq.~\eqref{eq:Delta_t_dec_NLO_heuristic} make the FAR smaller than the upper bound of Eq.~\eqref{eq:FARtot_approx}. Furthermore, we find that in this model the magnitude of the corrections is governed by $\rNLO$, Eq.~\eqref{eq:rho_NLO}, which is a parameter that characterizes how the correlation $|\Gamma(\Delta t)|$ deviates from a parabola around $\Delta t = 0$. From Eq.~\eqref{eq:Delta_t_dec_NLO_heuristic} we observe that when we increase the SNR threshold  $\rho$, the magnitude of the correction decays as $(\rNLO/\rho)^2$, and so for $\rho \gtrsim 3 \rNLO$, the relative error done when ignoring these corrections is smaller than $\sim 10\%$.

\begin{figure}[t!]
\begin{center}
\includegraphics[width=0.48\textwidth]{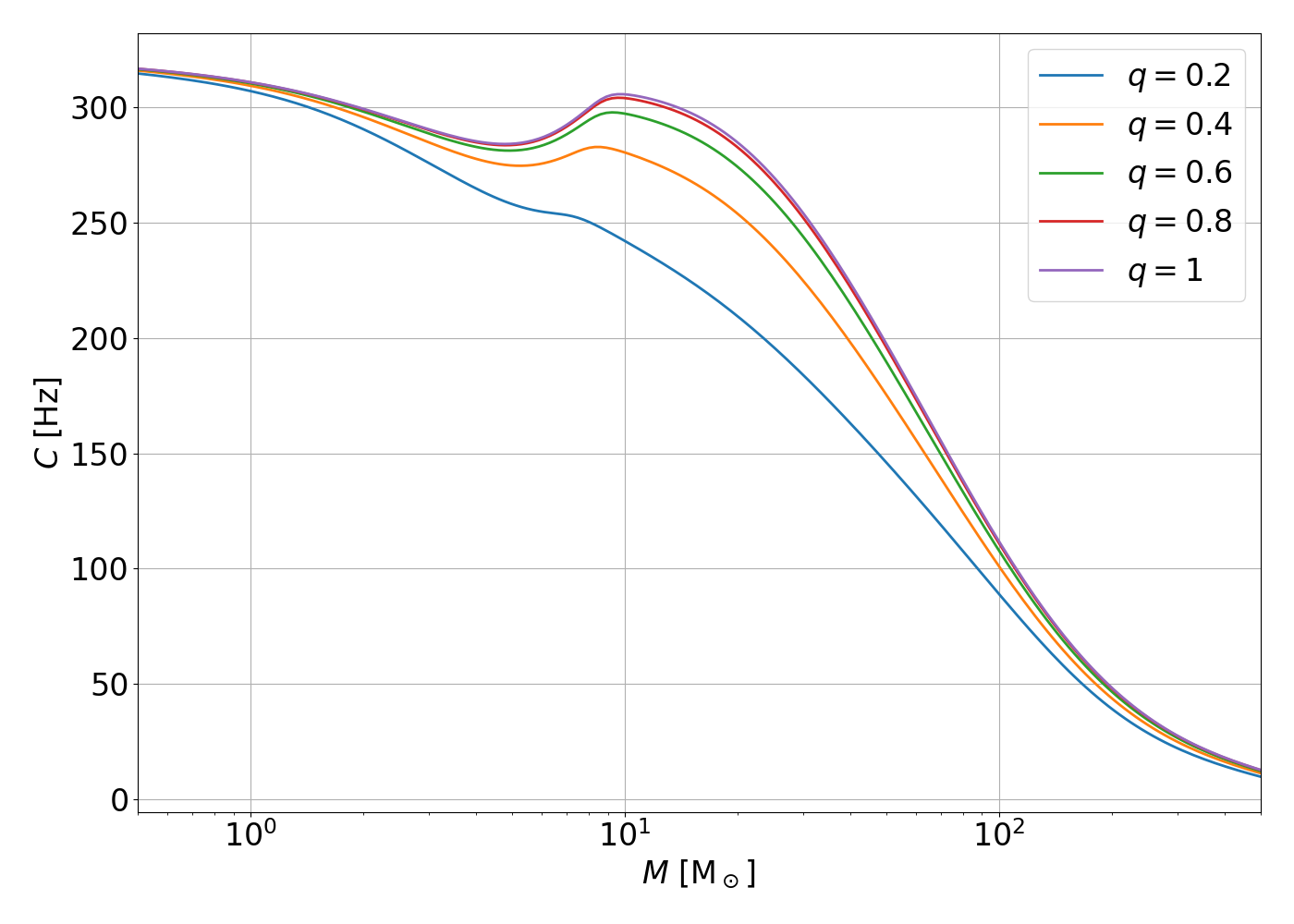}
\end{center} 
\caption{The FAR prefactor $C$ as a function of the CBC masses parameterized via the total mass of the binary $M=m_1+m_2$ and the mass ratio $q=m_2/m_1$ and computed using the PSD of Advanced LIGO at design sensitivity \cite{AdvLIGO_design} between $f_\mathrm{min} = 20$~Hz and $f_\mathrm{max} = 2048$~Hz. The waveform has been computed using \texttt{IMRPhenomPv2} with zero spin.}
\label{fig:coef_Mtot}
\end{figure}

\begin{figure}[t!]
\begin{center}
\includegraphics[width=0.48\textwidth]{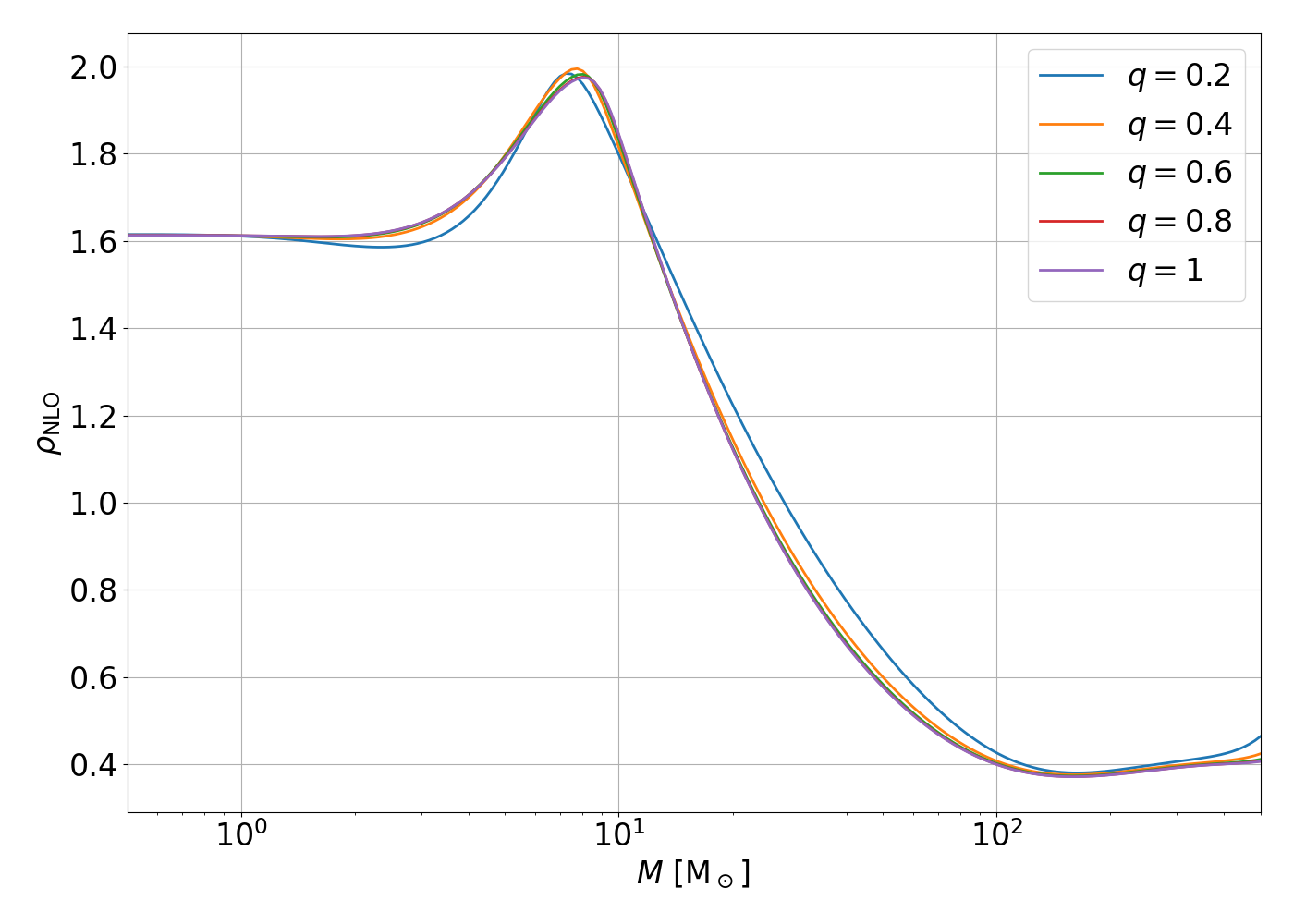}
\end{center}
\caption{The $\rNLO$ as a function of the CBC masses parameterized via the total mass of the binary $M=m_1+m_2$ and the mass ratio $q=m_2/m_1$ and computed using the PSD of Advanced LIGO at design sensitivity \cite{AdvLIGO_design} between $f_\mathrm{min} = 20$~Hz and $f_\mathrm{max} = 2048$~Hz. The waveform has been computed using \texttt{IMRPhenomPv2} with zero spin.}
\label{fig:snrNLO_Mtot}
\end{figure}

\subsection{Dependence on the CBC template parameters}
At a constant matched filter SNR, and neglecting higher order corrections ($\rho \gg \rNLO$), the False Alarm Rate of Eq.~\eqref{eq:FARtot_approx} will only depend on the signal via the multiplicative coefficient $C$ defined in Eq.~\eqref{eq:C_coef}, which when multiplied by the SNR, gives us the effective sampling rate. Since the higher the effective sampling rate, the more false alarms we expect, we can study how much Gaussian noise background there is in different regions of the CBC parameter space by representing the coefficient $C$ as a function of the CBC parameters. This is done in the Fig.~\ref{fig:coef_Mtot}, where we plot $C$ as a function of the CBC component masses for the \texttt{IMRPhenomPv2} waveform with the spins set to 0. 

The masses are parameterized via the total mass of the binary, $M=m_1+m_2$, and the mass ratio, $q=m_2/m_1$, the leading order parameters that control the amplitude evolution of the waveform \cite{IMR_approx}, which is the part that enters in the computation of $C_k$ in Eq.~\eqref{eq:C_k}. These results are robust with respect to the waveform choice since $C_k$ depends only on the amplitude evolution which is not as sensitive to modeling uncertainties as quantities that depend on the phase evolution of the template~\cite{TheLastThreeMinutes}.

In Fig.~\ref{fig:coef_Mtot} we obtain the natural result that, as a general trend, the higher the mass, the smaller the FAR will be (at a constant $\rho$). This is because the characteristic frequency of the event will be smaller, and then the characteristic autocorrelation time of the matched filter SNR will be longer, meaning that the time between independent trials will be longer. On top of this general trend we observe a peak at around $M \sim 10 M_\odot$, which will be due to events whose merger lies in the upper part of the most sensitive frequency range of the interferometer. Since during merger $|\tilde{h}(f)|^2\propto f^{-4/3}$ instead of $|\tilde{h}(f)|^2\propto f^{-7/3}$ as in the inspiral \cite{IMR_approx}, this will make $g(f)$ (Eq.~\eqref{eq:eff_prob_dist}) decay slower at larger frequencies where it is usually suppressed by the quantum shot noise ($S_n(f) \propto f^2$ \cite{Kimble_2001} at high frequency). In this case where merger lies in the upper part of the most sensitive frequency range of the interferometer, the value of $C$ will be larger because the band of frequencies that contribute will be larger. As a consequence of $C$ being larger, the effective sampling rate will be larger, leading to more false alarms. 

In Fig.~\ref{fig:snrNLO_Mtot}, the parameter $\rNLO$ giving the scale of the next to leading order corrections is shown. This quantity has a similar behavior as that of $C$, saturating at small masses where the merger is outside the sensitivity band, and generally decreasing at large masses whose merger happens at low frequency. It also has a peak at intermediate masses, corresponding to those systems that merge in the upper range of the frequency band that has the highest sensitivity. Note that in the case of $\rNLO$, this peak is more pronounced and towards smaller masses than in the case of $C$, which is due to the fact that in this range the value of $\rNLO$ is dominated by the value of $C_4$, which weighs more heavily higher frequencies than $C_2$, see Eq.~\eqref{eq:C_k}. The maximum of $\rNLO$ is achieved in this peak around $M \sim 8 M_\odot$, with a value of $\rho_\mathrm{NLO, max} \sim 2$. This means that if we go to $\rho \gtrsim 6$, the relative magnitude of the deviations from Eq.~\eqref{eq:FARtot_approx} will be smaller than $\sim 10\%$ for all CBC parameter range (see Eq.~\eqref{eq:Delta_t_dec_NLO_heuristic}). Therefore, as long as $\rho \gtrsim 6$ Eq.~\eqref{eq:FARtot_approx} will not only be an upper bound, but also a very good approximation of the FAR.

\begin{figure}[t!]
\begin{center}
\includegraphics[width=0.5\textwidth]{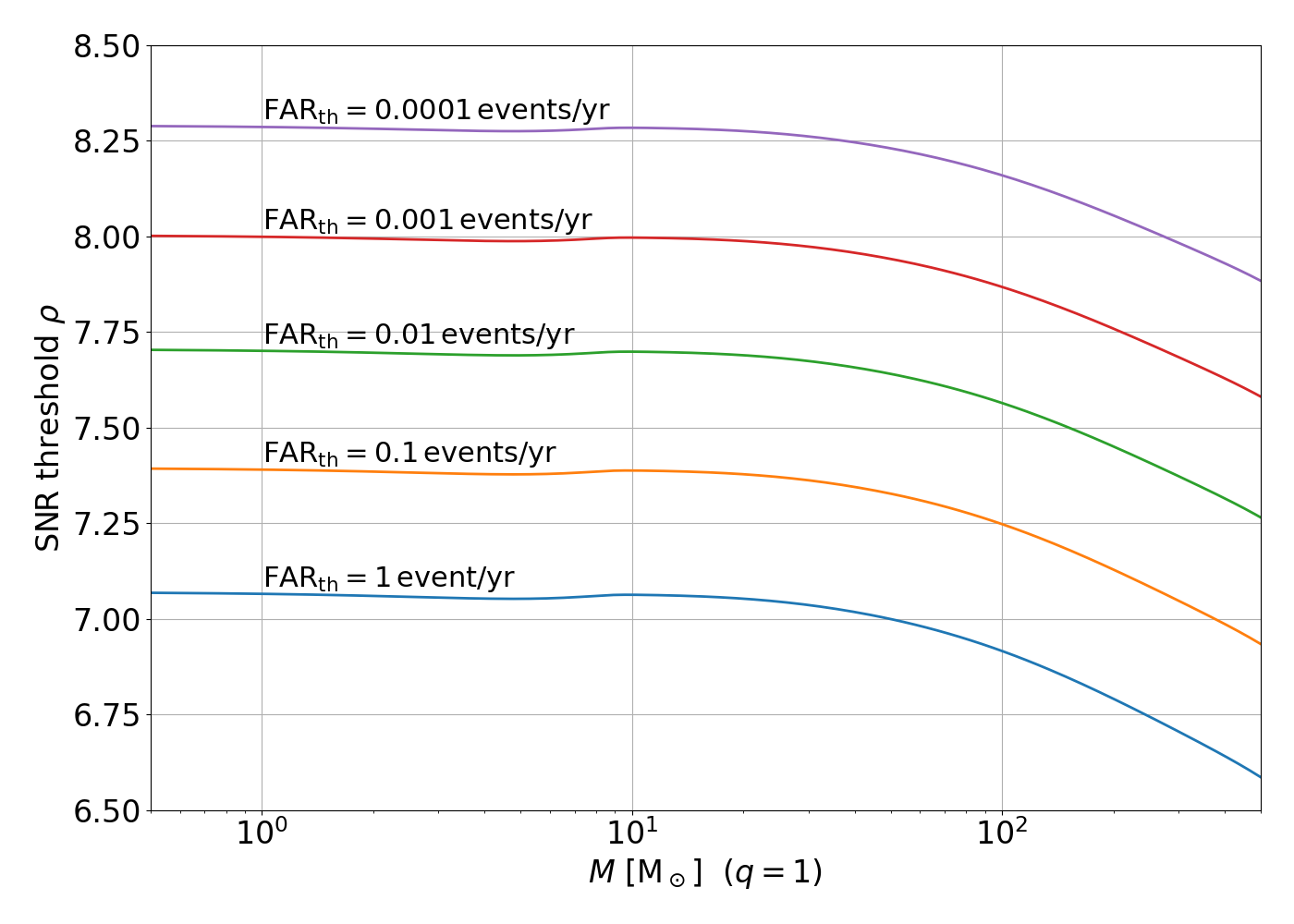}
\end{center} 
\caption{SNR threshold $\rho$ for different FAR thresholds as a function of the total mass of the binary $M$, assuming 0 spin and equal component masses ($q=1$). $\rho$ is plotted using Eq.~\eqref{eq:snr_th_FAR_th}, where the values of $C$ are the same as the ones for the $q=1$ curve of the left panel of Fig.~\ref{fig:coef_Mtot}. Direct comparison between the approximation of Eq.~\eqref{eq:snr_th_FAR_th} and the $\rho$ obtained by numerically inverting Eq.~\eqref{eq:FARtot_approx} shows that the maximum relative error made on $\rho$ is of $2\times 10^{-5}$ for the values shown in this plot.
}
\label{fig:snr_thr_Mtot}
\end{figure}

Having established the validity of Eq.~\eqref{eq:FARtot_approx} to approximate the FAR, we can now use it to find what SNR threshold $\rho$ would we need to set to discard all events with FAR higher than a given threshold FAR$_\mathrm{th}$. To do this we have to invert Eq.~\eqref{eq:FARtot_approx}, which can not be done exactly in terms of elementary functions, since it is a transcendental equation, but it can be done approximately in the limit that $\rho \gg 1$: 
\begin{equation}
    \rho =\! \sqrt{2 \log\frac{C}{\mathrm{FAR}_\mathrm{th}} + \log\!\left\{\!2\log\frac{C}{\mathrm{FAR}_\mathrm{th}}\!\right\}\!\! \left(\! 1 + \frac{1}{2 \log\frac{C}{\mathrm{FAR}_\mathrm{th}}}\!\right)} ,
    \label{eq:snr_th_FAR_th}
\end{equation}

\noindent which gives $\rho$ with a relative error of order $O(\log^2(\rho)/\rho^6)$. 
In Fig.~\ref{fig:snr_thr_Mtot} we have plotted this SNR threshold $\rho$  for different FAR thresholds as a function of the total mass of the binary $M$, assuming equal component masses ($q=1$). Even though from Fig.~\ref{fig:coef_Mtot} we observe that the value of $C$ depends strongly on $M$, when we introduce this $C$ in Eq.~\eqref{eq:snr_th_FAR_th}, $\rho$ depends to leading order on the square root of its logarithm and so has only a mild dependence on $M$ as can be seen in Fig.~\ref{fig:snr_thr_Mtot}. As a general trend, the higher $M$ is, the smaller the SNR threshold $\rho$ will have to be set to exclude false alarms at a given rate $\mathrm{FAR}_\mathrm{th}$, with the peak at $M \sim 10 M_\odot$ that was was observed in Fig.~\ref{fig:coef_Mtot} now less prominent due to the logarithmic dependence. The dependence on $\mathrm{FAR}_\mathrm{th}$ will also be mild, as $\rho$ will also depend on the square root of the logarithm of this quantity. Because of this, the variation of an order of magnitude in $\mathrm{FAR}_\mathrm{th}$ changes $\rho$ by only a small amount. We observe that if we set $\rho = 8$, as is commonly done in the theoretical literature \cite{Chen:2017wpg}, we would be rejecting Gaussian noise false alarms with rates higher than $\mathrm{FAR}_\mathrm{th} \sim 10^{-3} \mathrm{yr}^{-1}$.

\section{The false alarm rate of a template in a network of detectors}
\label{sec:SNR_FAR:ManyDet}

In this section we want to determine how much time of stationary Gaussian noise $n_i(t)$ of the detectors in a network would we have to look at on average to obtain a match with a signal to noise ratio (SNR) greater than some threshold $\rho$, using a GW template for the two polarization $\{h_+(t), h_\times(t)\}$, which when projected in the $i$-th detector leaves a signal $h_i(t)$. For the problem to be well-posed we will have first to define what we mean by the SNR for multiple detectors. In the case we have more than one detector, the total optimal SNR $\rho_\mathrm{tot}^\mathrm{opt}$ is defined by summing the individual optimal SNRs (Eq.~\eqref{eq:SNR_opt_def}) in quadrature, that is:

\begin{equation}
    \rho_\mathrm{tot}^\mathrm{opt} = \sqrt{\sum_i \langle h_i , h_i \rangle_i} = \sqrt{\sum_i (\rho_i^\mathrm{opt})^2 }   \, ,
    \label{eq:SNR_opt_ManyDet_def}
\end{equation}

\noindent where $\langle \times, \times \rangle_i$ denotes the inner product (Eq.~\eqref{eq:inner_product}) with the PSD $S_{i}(f)$ of the $i$-th detector. If $s_i(t)$ is the strain data in the $i$-th detector of the network, then the total matched filter SNR $\rho_\mathrm{tot}^\mathrm{mf}$ is defined as:

\begin{equation}
    \rho_\mathrm{tot}^\mathrm{mf} = \frac{1}{\rho_\mathrm{tot}^\mathrm{opt}} \sum_i \langle h_i , s_i \rangle_i  = \frac{1}{\rho_\mathrm{tot}^\mathrm{opt}} \sum_i \rho_i^\mathrm{opt} \rho_i^\mathrm{mf} \, ,
    \label{eq:SNR_mf_ManyDet}
\end{equation}

\noindent which given that each $\rho_i^\mathrm{mf}$ is a complex normal variable, if there are no correlations between detectors, will also be a complex normal variable. As was the case for the single detector matched filter SNR, the real part of Eq.~\eqref{eq:SNR_mf_ManyDet} will be the optimal quantity to rank the triggers when the form of the signal is known. Nonetheless, as was discussed in Sec.~\ref{sec:SNR_FAR}, in most cases of interest, 
the global phase of the GW can be changed arbitrarily and does not carry any information. 
Therefore we want to set $|\rho_\mathrm{tot}^\mathrm{mf}|$ as the ranking statistic, so that we get rid of the global phase while keeping the information contained in the relative phase and time of arrival of the GW in each detector, which will be related to the orientation and location of the detectors with respect to the direction and orientation of the GW source.
The relative phase of the incoming GW in the different detectors is sometimes ignored in GW searches to reduce computational cost and can easily add single detector triggers~\cite{PyCBC_2016}, although methods to take it into account in a statistical way have recently been introduced~\cite{Nitz:2017svb}. The relative phases between detectors are ignored when using the incoherent SNR, which is obtained adding the absolute value of the single detector matched filter SNRs in quadrature:

\begin{equation}
    \rho^\mathrm{inc} = \sqrt{\sum_i |\rho_i^\mathrm{mf}|^2} \, .
    \label{eq:SNR_incoh}
\end{equation}

\begin{figure*}[t!]
\begin{center}
\includegraphics[width=\textwidth]{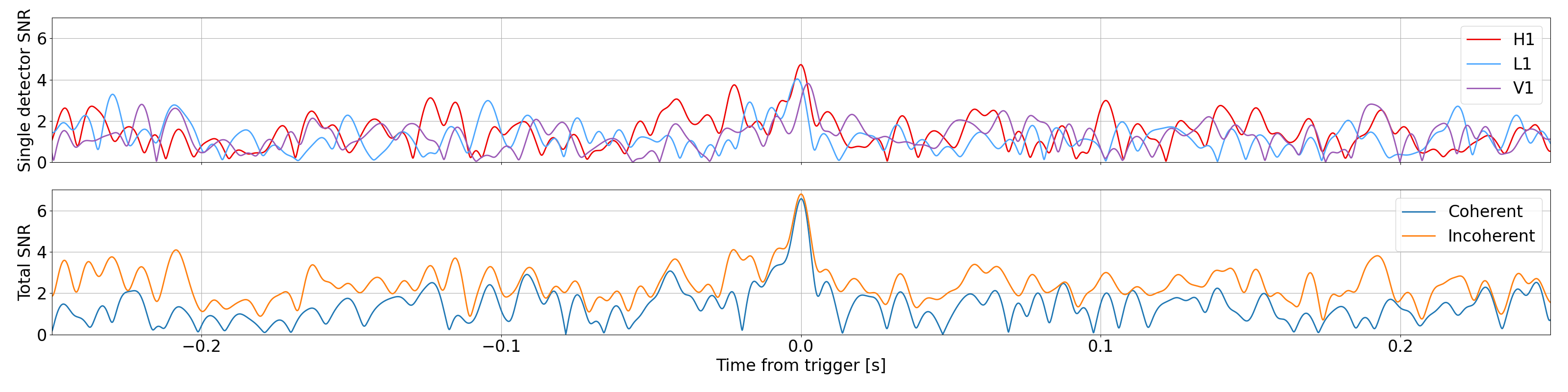}
\end{center} 
\caption{Top panel: Simulation of the modulus of the single detector matched filter SNR $|\rho^\mathrm{mf}(t)|$ for Gaussian noise generated of H1, L1 and V1 at design sensitivity. The match is performed using a spinless \texttt{IMRPhenomPv2} \cite{Khan_2019} template of masses $m_1 = m_2 = 50 M_\odot$, with extrinsic parameters right ascension 1.7rad, declination 1.7rad, polarization 0.2rad and a reference time $t_\mathrm{GPS} = 1000000000s$. Bottom Panel: We show the result of adding the single detector SNRs of the top panel both coherently (Eq.~\eqref{eq:SNR_mf_ManyDet}) and incoherently (Eq.~\eqref{eq:SNR_incoh}). To obtain the particular curves shown in this plot we generated random Gaussian noise in the three interferometers until we obtained a realization with a time at which $|\rho_\mathrm{tot}^\mathrm{mf}|>6.5$ and we plot 0.5s around the maximum of this trigger.
}
\label{fig:Gaussian_snr_t_realization_manydet}
\end{figure*}

Nonetheless, in this paper this ranking statistic will not be used as a lot of information is lost with it. If we shift in time the signals in all detectors, they will change by the same factor ($\mathcal{F}(h_i(t')) = \mathcal{F}(h_i(t)) e^{-2\pi i f (t'-t)}$), and then as in Eq.~\eqref{eq:SNR_mf_t} we can compute the matched filter SNR of the signal at different times with Gaussian noise using the following expression:

\begin{equation}
        \rho_\mathrm{tot}^\mathrm{mf}(t) = \frac{4}{\rho_\mathrm{tot}^\mathrm{opt}}  \int_{f_\mathrm{min}}^{f_\mathrm{max}} d f e^{2\pi i f t} \sum_i \frac{\tilde{h}_i^{*}(f) \tilde{n}_i(f)}{S_i (f)}  \, ,
    \label{eq:SNR_mf_ManyDet_t}
\end{equation}

\noindent where $S_i (f)$ is the noise PSD in the $i$-th detector. This quantity will also have correlations between different times that will affect the false alarm rate in a very similar way as in Sec.~\ref{sec:SNR_FAR}. This correlation can be explicitly seen in Fig.~\ref{fig:Gaussian_snr_t_realization_manydet}, where in the top panel we have plotted a simulation similar to that of Fig.~\ref{fig:Gaussian_snr_t_realization} for a random realization of the matched filter SNR for each detector in a Network formed by LIGO Livingston (L1), LIGO Hanford (H1) \cite{AdvLIGO_design} and Virgo (V1) \cite{AdvVirgo_design} at their design sensitivities. In the bottom panel we plot the sum of these single detector SNRs both in a coherent way (Eq.~\eqref{eq:SNR_mf_ManyDet}) and incoherent way (Eq.~\eqref{eq:SNR_incoh}). We observe how these two are smooth functions and are thus autocorrelated in time. We also observe that the incoherent SNR is always above the coherent one (sometimes quite significantly), since it ignores the important information carried by the consistency of the GW phase in the different detectors.

In a similar way as in Eq.~\eqref{eq:SNR_mf_t_autocorr}, we can quantify the autocorrelation in time of $\rho_\mathrm{tot}^\mathrm{mf}(t)$ by computing the covariance between different times:

\begin{widetext}
\begin{align}
    \Gamma(t,t')  & = \frac{1}{2}\langle \rho_\mathrm{tot}^\mathrm{mf}(t) \rho_\mathrm{tot}^\mathrm{mf}(t')^{*}\rangle =  \nonumber \\
    & = \frac{8}{(\rho_\mathrm{tot}^\mathrm{opt})^2} \left \langle  \! \int_{f_\mathrm{min}}^{f_\mathrm{max}} \! \! \! d f \int_{f_\mathrm{min}}^{f_\mathrm{max}} \! \! \! d f' e^{2\pi i (f t - f' t')} \sum_i \sum_j \frac{\tilde{h}_i^{*}(f) \tilde{h}_j(f') \tilde{n}_j^{*}(f') \tilde{n}_i(f)}{S_i (f) S_j(f')}  \! \right \rangle \nonumber \\
    & = \frac{8}{(\rho_\mathrm{tot}^\mathrm{opt})^2} \int_{f_\mathrm{min}}^{f_\mathrm{max}} \! \! \! d f \int_{f_\mathrm{min}}^{f_\mathrm{max}} \! \! \! d f' e^{2\pi i (f t - f' t')} \sum_i \sum_j \frac{\tilde{h}_i^{*}(f) \tilde{h}_j(f') \langle \tilde{n}_j^{*}(f') \tilde{n}_i(f)\rangle}{S_i (f) S_j(f')}  \nonumber \\
    & = \Gamma(t-t') = \frac{4}{(\rho_\mathrm{tot}^\mathrm{opt})^2} \int_{f_\mathrm{min}}^{f_\mathrm{max}} \! \! \! d f e^{2\pi i f (t-t')} \sum_i  \frac{|\tilde{h}_i(f)|^2}{S_i (f)}  \, , \nonumber \\
    \label{eq:SNR_mf_t_autocorr_ManyDet}
\end{align}
\end{widetext}

\noindent where we have used that when there is no correlation between the noise of different detectors, then $\langle \tilde{n}_j^{*}(f') \tilde{n}_i(f)\rangle = \frac{1}{2} S_i (f) \delta_{i j} \delta(f-f')$. What we observe in Eq.~\eqref{eq:SNR_mf_t_autocorr_ManyDet} is that in the many detector case we obtain the same formula of the covariance as in the single detector case of Eq.~\eqref{eq:SNR_mf_t_autocorr} if we do the following identification

\begin{equation}
    \frac{1}{(\rho^\mathrm{opt})^2} \frac{|\tilde{h}(f)|^2}{S_n (f)} \longrightarrow  \frac{1}{(\rho_\mathrm{tot}^\mathrm{opt})^2} \sum_i  \frac{|\tilde{h}_i(f)|^2}{S_i (f)}\,.
    \label{eq:Onedet_to_Manydet}
\end{equation}

Therefore the FAR will be given by the same expressions that were found in Sec.~\ref{sec:SNR_FAR} for the single detector case doing the identification of Eq.~\eqref{eq:Onedet_to_Manydet}. That is, an accurate upper bound approximation of the FAR is given by  Eq.~\eqref{eq:FARtot_approx}, with $C$ given by the same formula of Eq.~\eqref{eq:C_coef}, but now using the following expression for $C_k$: 

\begin{equation}
    C_k = \frac{4}{(\rho_\mathrm{tot}^\mathrm{opt})^2} \int_{f_\mathrm{min}}^{f_\mathrm{max}} \! df \, (2 \pi f)^k  \sum_i  \frac{|\tilde{h}_i(f)|^2}{S_i (f)}   \, .
    \label{eq:C_k_ManyDet}
\end{equation}

\section{Application to GW events}
\label{sec:SNR_FAR:GW_events}

So far we have discussed the FAR and the FAP for a predefined template given a threshold SNR $\rho$. However, in real settings what we observe is a fluctuation in the strain, that we do not know if it comes from a GW or from noise, and which we will generically call an event. This fluctuation can be interpreted under any template, each giving a different SNR. For a given template, the threshold SNR $\rho$ to use in Eq.~\eqref{eq:FARtot_approx} for the FAR computation is the observed total matched filter SNR ($\rho = |\rho_\mathrm{tot}^\mathrm{mf}|$), since we want to know how likely it is to find SNRs equal to or larger than the one observed for the template. The problem will then be how to choose a template, given the observed strain, to determine the SNR and to compute the FAR using Eq.~\eqref{eq:FARtot_approx}. The likelihood is the conditional probability of obtaining the observed strain given a GW signal with parameters $\vec{\theta}$. If we assume Gaussian noise, the likelihood takes the the following form \cite{GWLikelihood_Finn}:

\begin{align}
    \mathcal{L}(\textbf{s} | \vec{\theta}) & = \mathcal{N} \exp\left\{-\frac{1}{2} \sum_i \langle s_i - h_i(\vec{\theta}), s_i - h_i(\vec{\theta}) \rangle_i \right\} \nonumber \\
    & \propto \exp\left\{ \rho_\mathrm{tot}^\text{opt}(\vec{\theta}) \! \left( \! \text{Re}\left\{\rho_\mathrm{tot}^\mathrm{mf}(\vec{\theta}, \textbf{s})\right\} - \frac{1}{2}\rho_\mathrm{tot}^\text{opt}(\vec{\theta}) \!\right)\! \right\} \,,
    \label{eq:Likelihood}
\end{align}
where $\mathcal{N}$ is a normalization constant. Note that the likelihood will be larger for those templates that have the largest matched filter SNR and an optimum SNR such that $\rho_\mathrm{tot}^\text{opt} = \text{Re}\{\rho_\mathrm{tot}^\mathrm{mf}\}$, which for GW templates can always be achieved by varying the distance to the source.  We then have the expected result that, the more SNR a template has, the larger its Likelihood is and, therefore, the more likely it is to reproduce the observed strain.

However, when we associate a template with an event, we are interpreting the strain fluctuation in terms of a model, with underlying assumptions about the possible physics. The consistent way to take this into account is to think of the event as having a probability of being described by any template, with some priors on each template \footnote{For example, even though the template that maximizes the SNR is the one that exactly reproduces the strain ($h_i(t) = s_i(t)$), this is usually a physically impossible GW template, and in this case, we will not consider it. We have that our prior probability for a template that can not be generated by GWs is 0.}. Because we are characterizing a fluctuation observed in the data, we need to evolve our priors to find the probability of each template describing the specific strain. Therefore, what naturally arises is the need to employ Bayes' Theorem to determine the posterior probability $p(\vec{\theta}|\textbf{s})$ of each template given the observed strain $\textbf{s}$:
\begin{equation}
    p(\vec{\theta}|\textbf{s}) = \frac{\mathcal{L}(\textbf{s} | \vec{\theta}) \pi(\vec{\theta})}{\int d \vec{\theta}' \mathcal{L}(\textbf{s} | \vec{\theta}') \pi(\vec{\theta}')}\,,
    \label{eq:Bayes_theorem}
\end{equation}
where $\pi(\vec{\theta})$ is the prior probability for each set of parameters and it is multiplied by the likelihood to give the posterior. The more SNR a template has, the larger its likelihood and the more weight it will be given in the posterior probability distribution. In Bayesian inference, the posterior $p(\vec{\theta}|\textbf{s})$ is interpreted as the probability of the template given the strain. Therefore, the template corresponding to the maximum of the posterior probability distribution is the most likely template given the strain and our priors, while the maximum likelihood template is the template most likely to generate the observed strain. In general, these two templates will be different from each other, and they will have different FAPs when computed with Eqs.~\eqref{eq:FAP_of_FAR},~\eqref{eq:FARtot_approx}, that we can call $\mathrm{FAP}_{\mathrm{max}\, p}$ and $\mathrm{FAP}_{\mathrm{max}\,\mathcal{L}}$ respectively. The most representative template when comparing to the LVK searches would correspond to the maximum likelihood sample, since the modeled searches performed by the LVK \cite{GWTC-3} deal with the unknown intrinsic parameters by setting up a template bank to cover a target parameter space, and then selecting the template which has the highest likelihood ratio for signal vs noise origin in a given segment of data which, in the Gaussian noise case, means the highest SNR sample. In practice, the FAR reported by LVK searches would be the FAR of this max likelihood template multiplied by the trial factor given by the number of independent templates within the search parameter space.

Another possibility to consider all the information contained in the posterior is to compute the FAP of the fluctuation. To do so, we combine the probability of each template describing the fluctuation given by the posterior, and the probability of each template to be generated by Gaussian noise with an SNR equal to or larger than the observed one, given by the FAP, see Eqs.~\eqref{eq:FAP_of_FAR},~\eqref{eq:FARtot_approx},
\begin{align}
    & \mathrm{FAP}_\mathrm{event}  = \int d \vec{\theta} p(\vec{\theta}|\textbf{s}) \mathrm{FAP}(\vec{\theta}, \textbf{s})  \nonumber \\ 
    & \!\! =\!\! \int \!\! d \vec{\theta} p(\vec{\theta}|\textbf{s}) \! \left(\! 1\! -\! \exp\!\left\{\!- T_\mathrm{obs} C(\vec{\theta}) \big|\rho_\mathrm{tot}^\mathrm{mf}(\vec{\theta}, \textbf{s})\big| e^{-\frac{1}{2}\big|\rho_\mathrm{tot}^\mathrm{mf}(\vec{\theta}, \textbf{s})\big|^2} \! \right\} \! \right) \! \,,
    \label{eq:FAP_event}
\end{align}
which will always be less than or equal to one, since the posterior $p(\vec{\theta}|\textbf{s})$ is normalized, as can be seen in Eq.~\eqref{eq:Bayes_theorem}. The FAP$_\mathrm{event}$ of Eq.~\eqref{eq:FAP_event} will now not only depend on a single template, but similarly to the Bayes Factor \cite{GW_BayesFactor} it will take into account the distribution of the likelihood over the prior volume. Therefore, it can be seen as an effective way of considering the trial factor for the template that best matches the data over a parameter space.

In general, the normalization of the posterior, given by the evidence $\mathcal{Z}\!=\!\int\! d \vec{\theta} \mathcal{L}(\textbf{s} | \vec{\theta}) \pi(\vec{\theta})$, is extremely difficult to compute. However, even though the full posterior is unknown, one can use Monte Carlo methods to obtain independent samples from it, as done in Parameter Estimation Analysis~\cite{IntroBayesInference_GWs}. In terms of these independent posterior samples, Eq.~\eqref{eq:FAP_event} can be approximated by:
\begin{equation}
        \mathrm{FAP}_\mathrm{event}=\frac{1}{N_s}\sum_{i=1}^{N_s} \mathrm{FAP}(\vec{\theta}_i, \textbf{s}) \, .
        \label{eq:FAP_event_MonteCarlo}
\end{equation}

\noindent where $N_s$ is the number of samples, and the error of approximating the integral by a sum over independent posterior samples is given by:

\begin{equation}
        \Delta \mathrm{FAP}_\mathrm{event}=\sqrt{\frac{1}{N_s (N_s - 1)}\sum_{i=1}^{N_s}\left( \mathrm{FAP}(\vec{\theta}_i, \textbf{s}) - \mathrm{FAP}_\mathrm{event}\right)^2} \, .
        \label{eq:error_FAP_event_MonteCarlo}
\end{equation}

\subsection{Application to GW candidates in GWTC-3}
\label{sec:SNR_FAR:GW_events:LVK}

\begin{figure}[t!]
\centering
\includegraphics[width=0.5\textwidth]{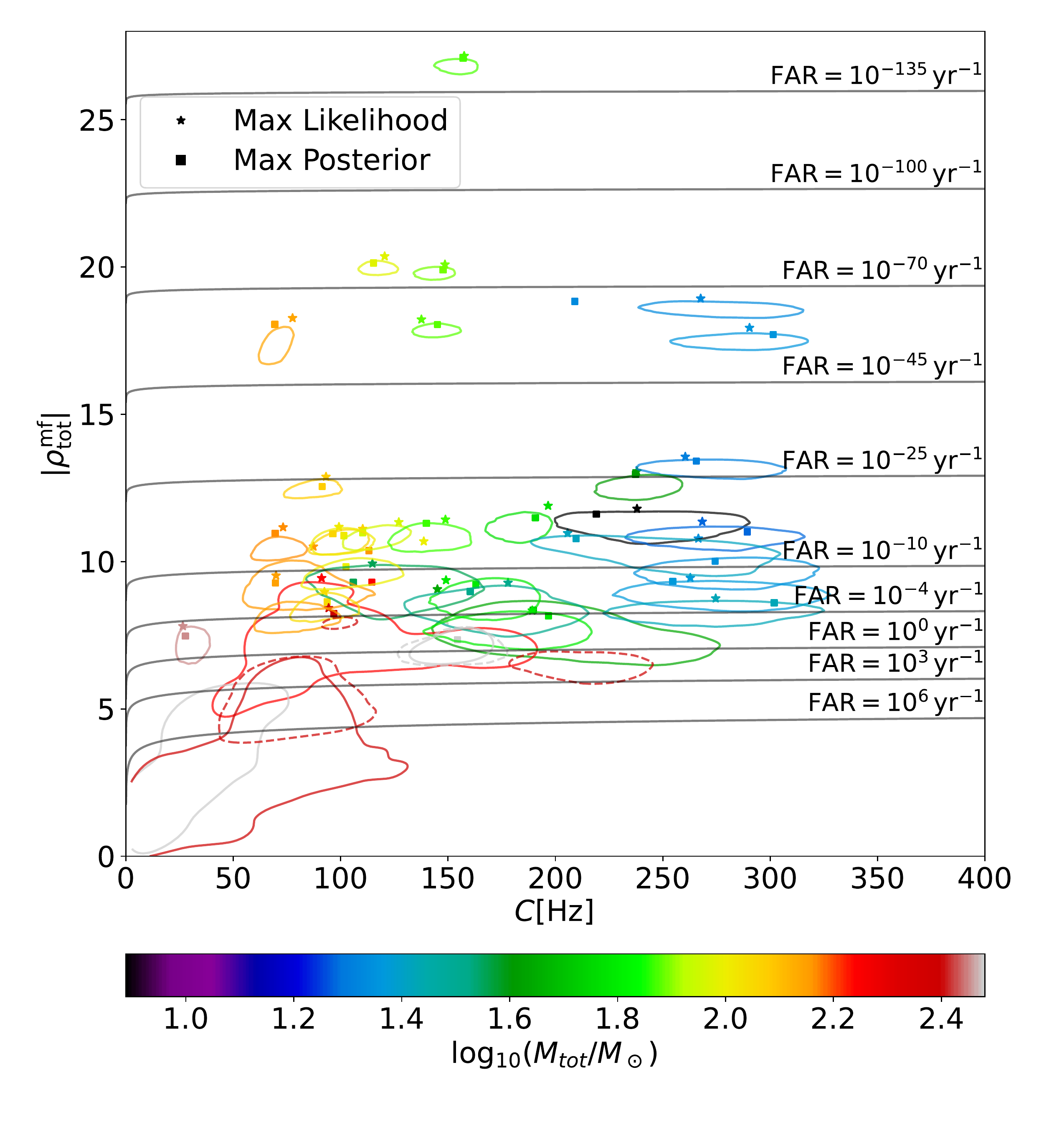}
\caption{In this plot we show the contours enclosing 90\% of the IMRPhenomXPHM~\cite{IMRPhenomXPHM} posterior samples in the ($C$, $|\rho^\mathrm{mf}_\mathrm{tot}|$) plane for all the O3b Catalog events. The value of $C$ is computed using Eqs.~\eqref{eq:C_coef},~\eqref{eq:C_k_ManyDet} with the local PSD around each event. The values of ($C$, $|\rho^\mathrm{mf}_\mathrm{tot}|$) for the maximum likelihood and maximum posterior probability samples are marked with a star and square respectively. We also plot using dashed lines the contours of the two events GW200308\_173609 (in grey) and GW200322\_091133 (in red) after making the cut in the Likelihood as was done for GWTC-3. The color of the contours is given by the median total mass of the posterior samples of each event. We also plot lines of constant FAR as defined by Eq.~\eqref{eq:snr_th_FAR_th}.}
\label{fig:rhoC_contours_GWTC3}
\end{figure}

As an application of the method previously outlined, we analyze the 35 CBC candidates included in the last gravitational wave transient catalog, GWTC-3 \cite{GWTC-3}, detected during the second part of the third observing run (O3b). The Bayesian Parameter Estimation (PE) of these events has been performed by the LVK collaboration as described in Ref.~\cite{GWTC-3} and the posterior samples obtained are publicly available in Ref.~\cite{PE_samples_GWTC-3}. 

We compute the value of $C$ for each IMRPhenomXPHM~\cite{IMRPhenomXPHM} sample of every event in GWTC-3, using Eqs.~\eqref{eq:C_coef},~\eqref{eq:C_k_ManyDet}, where we use the local PSD around each event that is the same one employed in the PE, also available in Ref.~\cite{PE_samples_GWTC-3}. In Fig.~\ref{fig:rhoC_contours_GWTC3} we show the 90\% credible intervals of $|\rho^\mathrm{mf}_\mathrm{tot}|$ and $C$, which are the contours enclosing 90\% of the posterior samples in the ($C$, $|\rho^\mathrm{mf}_\mathrm{tot}|$) plane. Since at first order the Gaussian FAR only depends on $C$ and $|\rho^\mathrm{mf}_\mathrm{tot}|$, we can plot on top of Fig.~\ref{fig:rhoC_contours_GWTC3} the contours of constant FAR using Eq.~\eqref{eq:snr_th_FAR_th}. We observe that for most of the events, almost all the samples are above a Gaussian FAR of 1 per year, meaning that we do not expect them to come from a Gaussian noise fluctuation. However, there are two notable exceptions which have almost no posterior support for templates with Gaussian FAR under 1 per year, which correspond to GW200308\_173609 (grey) and GW200322\_091133 (red), having only 4.16\% and 0.71\% of the posterior samples above this threshold respectively. These are the two events that were noticed in GWTC-3 to have multimodal posterior distributions, due to the likelihood not having a sufficiently large peak to dominate the posterior in all parameter space, which induces prior-dominated modes at large distances and high masses.

In GWTC-3, an ad hoc cut in the likelihood was made to get rid of these prior-dominated modes. For GW200308\_173609 the samples with $\log\{\mathcal{L}/\mathcal{L}_0\} < 10$ are removed while for GW200322\_091133 the samples with $\log\{\mathcal{L}/\mathcal{L}_0\} < 2$ are removed, where $\mathcal{L}_0 = \exp\left(-\sum_i \langle s_i, s_i \rangle/2\right)$ is the likelihood of the data given no signal, i.e. substituting $h = 0$ in Eq.~\eqref{eq:Likelihood}\cite{GW_BayesFactor}. We show with dashed lines the contour that encompasses in the ($C$,$|\rho^\mathrm{mf}_\mathrm{tot}|$) plane 90\% of the samples that remain after the ad hoc Likelihood cut. We observe that the result is to remove the lowest SNR samples (since the SNR and the Likelihood are intimately related) and it thus removes the posterior samples with the largest FAR. However, a large fraction of the remaining samples still have FARs larger than 1 per year, with 32.9\% and 96.8\% of them above this threshold for GW200308\_173609 and GW200322\_091133 respectively. 

Looking only at the maximum likelihood sample of these two events (marked with a star in Fig.~\ref{fig:rhoC_contours_GWTC3}), they have large SNR values of 8.00 for GW200308\_173609 and 8.42 for GW200322\_091133, which makes them have a single template FAR$_{\mathrm{max} \, \mathcal{L}}$ of 4.7$\times 10^{-4} \mathrm{yr}^{-1}$ and 9.9$\times 10^{-6} \mathrm{yr}^{-1}$ respectively, without taking into account any trial factor due to the fact that the likelihood is maximized over a parameter space.

The Gaussian FAR that we have presented here is not directly comparable with the FAR computed by the LVK search pipelines, since they differ in methodology in various ways. The search pipelines make use of a template bank and a different ranking statistic from the bare SNR to take into account the presence of non-Gaussianities. The ranking statistic assigned to each trigger by the pipelines is the one maximized over all the template bank covering the parameter space of the search, with the background estimated by doing time-shifts in detector data. Another difference is that pipelines do not coherently sum the signal from all interferometers, as this would not allow marginalizing over the location in the sky, polarization and neither to work with single detector triggers, making the search computationally cost prohibitive. For this same reason, the template bank of the searches often use simplified waveform models, ignoring effects such as precession or Higher Order modes and do a coarser sampling of the parameter space than what is done in a Parameter Estimation.

In table~\ref{table:Far_table} we present the most important parameters to quantify the significance of the events in GWTC-3, coming both from the LVK search and PE results and from our Gaussian FAR analysis. Looking at the rightmost column, we notice that there are several events with Gaussian FAPs (computed using Eq.~\eqref{eq:FAP_event_MonteCarlo} with $T_\mathrm{obs} = 1 \mathrm{yr}$) that are of order 1. The highest FAPs come, as expected, from  GW200308\_173609 and GW200322\_091133, which have FAPs of 0.97 and 0.99 respectively. After the Likelihood cut, the FAP of GW200308\_173609 improves substantially, becoming 0.44. However, that's not the case for GW200322\_091133, which keeps a very high FAP after the cut, with a value of 0.97 due to the fact that it has small SNR values in most of its posterior.

\begin{figure}[t!]
\includegraphics[width=0.5\textwidth]{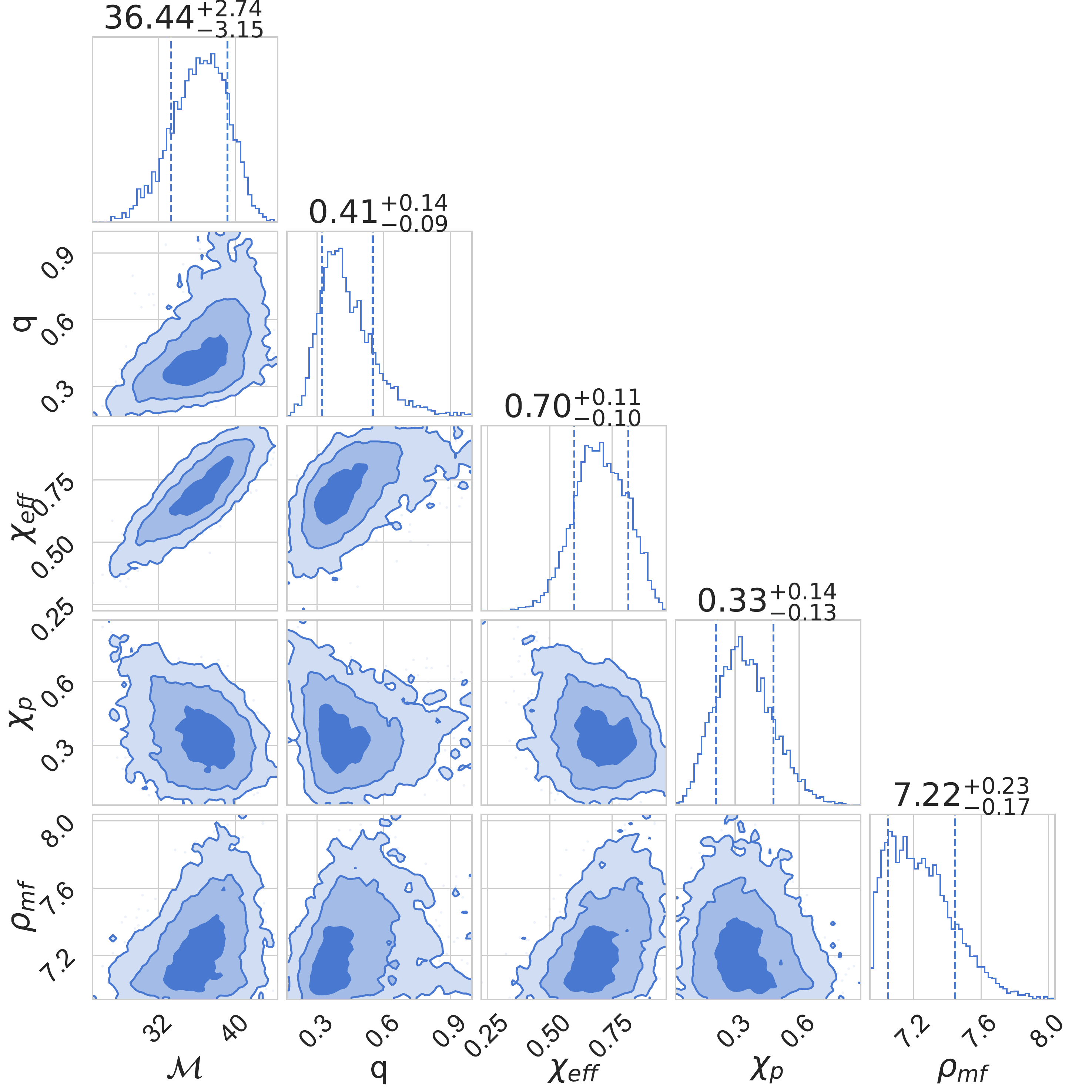} \\
\includegraphics[width=0.5\textwidth]{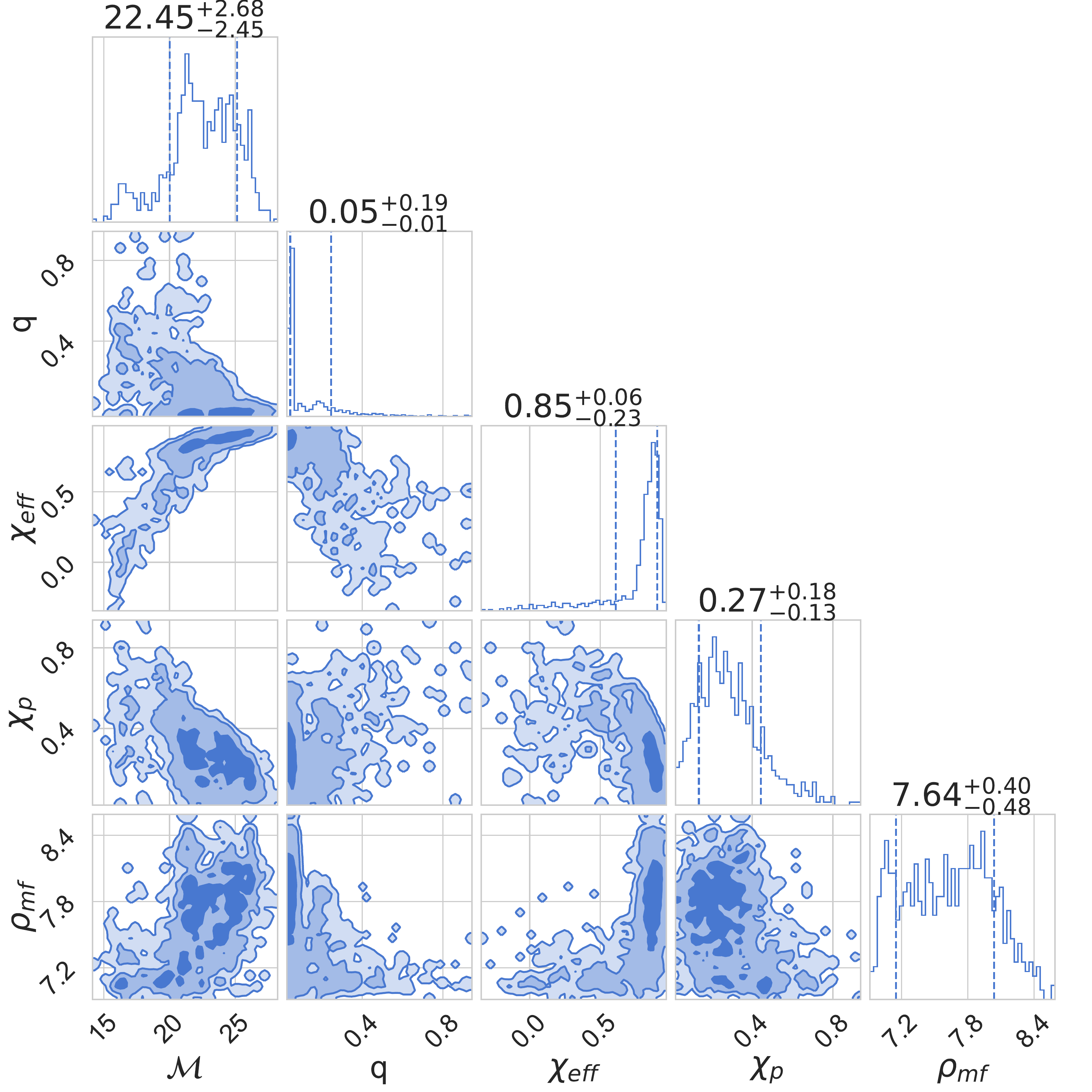}
\caption{Corner plots of selected parameters for the posterior samples with Gaussian FAR$\leq$1 per year. Top Panel: GW200308\_173609, Bottom Panel: GW200322\_091133}
\label{fig:BadEvents_FARcut}
\end{figure}

Since both GW200308\_173609 and GW200322\_091133 have a small subset of samples in their posteriors with larger SNRs and correspondingly small FARs, we can explore which samples have this larger significance by selecting only those that have a FAR below a $1\mathrm{yr}^{-1}$ threshold. In Fig.~\ref{fig:BadEvents_FARcut} we show the distribution of some of the binary parameters using only those samples with FAR below a $1\mathrm{yr}^{-1}$. We observe that the parameters of the waveforms that satisfy this cut are very different from all other CBC observations~\cite{GWTC3_population}, with both events having extremely large effective spin parameters $\chi_\mathrm{eff}$ and with GW200322\_091133 having a very extreme mass ratio for which waveform systematics might be important~\cite{IMRPhenomXPHM}. 
It's also noticeable that, due to the very low percentage of posterior samples with FAR below the $1{\rm yr}^{-1}$ threshold in GW200322\_091133, ($\sim0.07\%$), the parameter space might be undersampled.
In principle, both, the search \cite{Zenodo_Searches_Release} and the parameter estimation \cite{PE_samples_GWTC-3} should identify similar maximum likelihood points in the parameter space for a given trigger time. We can then compare the two template parameters' values as a sanity check. In the GW200308\_173609 case, differences in the masses are not significant, with trigger masses of $(m_1,m_2)=(58.4,41.3)M_\odot$ while the masses identified by the PE for the maximum likelihood template are $(m_1,m_2)=(64.2,38.2)M_\odot$. We find larger discrepancies for the GW200322\_091133, with trigger masses of $(m_1,m_2)=(56.0,15.3)M_\odot$ while the masses identified by the PE are $(m_1,m_2)=(161.3,7.8)M_\odot$. The calculation of $\pastro$ depends crucially on the values of the masses and such an extreme mass ratio would definitely represent an outlier to the population. For both events, in the search and in the maximum likelihood of the PE, very large values of $\chi_\mathrm{eff}$ are found, in contrast with the rest of the population of merging BH~\cite{GWTC3_population}. However, since the value of the spin is not taken into account for $\pastro$ calculations~\cite{MBTA_pastro_uncertinties}, this does not downrank the event. Finally, for the case of GW200322\_091133 we also find a substantial difference between the search SNR and the maximum likelihood SNR of the PE, being 9.0 and 8.4 respectively. Since the FAR and $\pastro$ have an exponential dependence with the SNR, this difference would also downweight the event.

\begin{figure}[t!]
\includegraphics[width=0.48\textwidth]{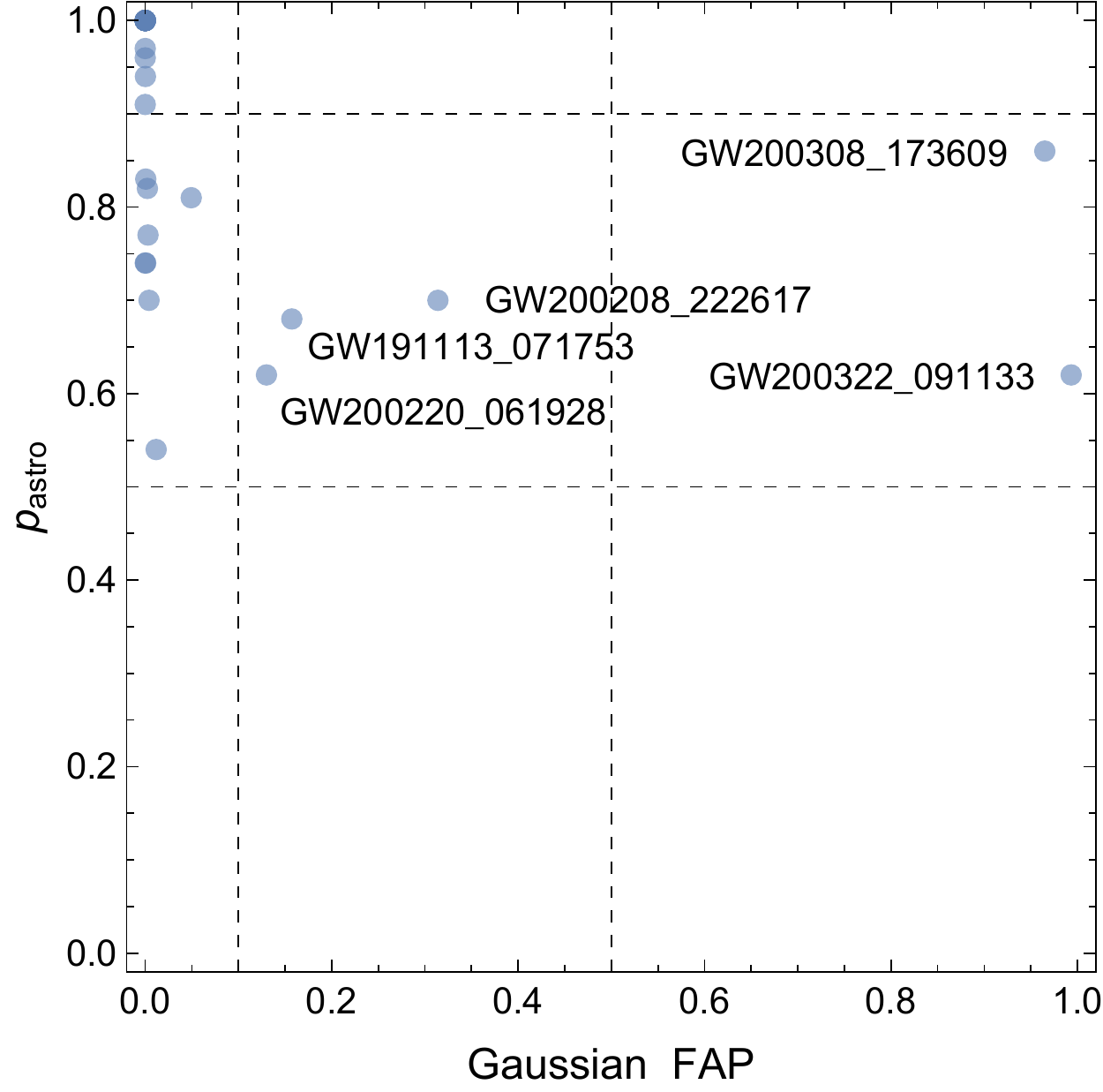}
\caption{The comparison in the Gaussian FAP and $\pastro$ plane of the different GWTC-3 events from Table~\ref{table:Far_table}. We have set cuts in FAP of 10\% and 50\%, as well as in $\pastro=0.5$ and 0.9.}
\label{fig:FAPpastro}
\end{figure}

Since our method only gives a lower bound estimation on the FAP, it does not allow us to state that a candidate is indeed a gravitational wave event, but it can support the hypothesis of a noise origin. We can derive how likely Gaussian noise is to generate a signal, but we can not say anything about the possibility of non-Gaussianities mimicking it.
With this in mind, in Fig.~\ref{fig:FAPpastro} we show how the GWTC-3 events are distributed in the Gaussian FAP and $\pastro$ plane (the values are taken from Table~\ref{table:Far_table}). We note that for all the 22 events with $\pastro>0.9$, the Gaussian FAP also gives them low probability of generation from a Gaussian noise fluctuation, having all $\mathrm{FAP} \leq 2\times10^{-4}$ and there is no inconsistency.
However, for the 13 events with $0.5<\pastro<0.9$, results are mixed. The majority of these events (8/13) also have Gaussian FAP smaller than 10\% and so we find that they are not likely to be generated from a Gaussian noise fluctuation. From the 13 events with $0.5<\pastro<0.9$ we have another 3 in the region of $10\% < \mathrm{FAP} < 50\%$, which therefore have some non-negligible probability of being generated by Gaussian noise, although it is still more likely they are not. These 3 events correspond to GW191113\_071753, GW200208\_222617 and GW200220\_061928, from which GW200208\_222617 is the one with the largest Gaussian FAP ($\sim 31\%$) and also has a multimodal posterior distribution~\cite{GWTC-3}. Finally, at $0.5<\pastro<0.9$ and FAP$> 50\%$ we have 2 points corresponding to GW200308\_173609 and GW200322\_091133 and which have already been discussed in detail as likely to be generated by a Gaussian fluctuation. It is also interesting to note that for all events with FAP $>$ 0.1, the $\pastro$ value quoted in GWTC-3~\cite{GWTC-3} is larger than 0.5 in only one of the pipelines, the others quoting significantly lower values.

\begin{Far_table}
    \begin{table*}
    	\input{Far_table}
    	\caption{In this table we report every candidate GW signal included in the O3b Catalog, as well as the detectors observing at the merger time of the events, the search pipeline in which it had the highest $p_{\rm astro}$ together with the Search estimated SNR, the Search FAR and the $p_{\rm astro}$ as calculated by that same pipeline. We also include the SNR as obtained by the LVK parameter estimation analysis, our Gaussian FAR and Gaussian FAP of the event assuming an observing time of 1yr. While the errors on the PE SNR and the Gaussian FAR represent the 90\% credible intervals, for the Gaussian FAP they represent the uncertainty on the Monte Carlo integral used to compute it, given by Eq.~\eqref{eq:error_FAP_event_MonteCarlo}. The events that have an asterisk and are in \textit{italic}, correspond to the ones in which we have performed the ad hoc cut in the Likelihood. 
    	}
    	\label{table:Far_table}
    \end{table*}
\end{Far_table}

\section{Conclusions}
\label{sec:Conclusions}

Understanding whether triggers in LIGO-Virgo detectors are from gravitational wave or noise origin is a hard task. For most of the events, the GW signal is expected to be extremely weak and in this paper we have explored the possibility of it being mimicked by the irreducible Gaussian noise in the gravitational wave detectors. 

We have derived a mathematical framework for estimating the rate of false alarms induced by this Gaussian noise. Our main result is given in Eq.~\eqref{eq:FARtot_approx}, which gives the rate at which the matched filter SNR of a specific template with the Gaussian noise of one (or multiple) GW detectors goes over a threshold $\rho$. The prefactor $C$ multiplying the FAR depends on the specific template used for matched filtering. For CBC templates the most important parameter controlling the value of $C$ is the total mass of the event, with $C$ being significantly smaller for larger masses. 

We have then studied how the Gaussian FAR of CBC templates behaves as a function of the threshold SNR, and gave an analytical expression for the minimum SNR needed for a given FAR threshold.
We have also proposed a method to estimate the probability of Gaussian noise with the local PSD mimicking a given GW candidate in terms of a false alarm probability~\eqref{eq:FAP_event}, using the samples from the Parameter Estimation analysis of such an event.

Finally, we have applied this formalism to the GW candidates that were added in the GWTC-3 catalog, obtaining a Gaussian FAR for each template in their PE posterior and a FAP for the events.

Summarizing these results, most of the samples of the events are clearly above the 1 $\mathrm{yr}^{-1}$ FAR threshold with event FAPs ranging from $\sim 10^{-143}$ to a more modest $\sim 10^{-1}$, assuming a reference observation time of one year. However, we find two clear outliers, GW200308\_173609 and GW200322\_091133, with event FAPs very close to one, signaling very high odds of Gaussian noise fluctuations mimicking them. We also explore the samples in their posterior that have single template FAR$<1 \,\mathrm{yr}^{-1}$. These samples have very extreme parameter values with respect to the observed BBH population, and in the case of GW200322\_091133 differ from those identified by the search.

We believe that the methods developed here may be useful in the future to further investigate GW triggers that are found in future LVK runs.

\section*{Acknowledgements}

The authors thank Thomas Dent and Viola Sordini for their helpful comments and discussions as reviewers of this paper in LIGO and Virgo respectively. The authors acknowledge use of the publicly available codes: \texttt{lalsuite} \cite{lalsuite_code}, \texttt{Bilby} \cite{Bilby_code}, \texttt{PyCBC} \cite{PyCBC_code}. 
They acknowledge support from the research project  PGC2018-094773-B-C32, and the Centro de Excelencia Severo Ochoa Program CEX2020-001007-S, while  
GM acknowledges support from the Ministerio de Universidades through Grant No. FPU20/02857
and JFNS acknowledges support from MCIN through Grant No. PRE2020-092571.
ERM is grateful to the Instituto de F\'isica Te\'orica (IFT) for their hospitality.
The authors acknowledge use of the Hydra cluster at the IFT, on which some of the numerical computations for this paper took place.
The authors are grateful for computational resources provided by the LIGO Laboratory and supported by National Science Foundation Grants PHY-0757058 and PHY-0823459.
This research has made use of data or software obtained from the Gravitational Wave Open Science Center \cite{LIGOScientific:2019lzm} (gw-openscience.org), a service of LIGO Laboratory, the LIGO Scientific Collaboration, the Virgo Collaboration, and KAGRA. LIGO Laboratory and Advanced LIGO are funded by the United States National Science Foundation (NSF) as well as the Science and Technology Facilities Council (STFC) of the United Kingdom, the Max-Planck-Society (MPS), and the State of Niedersachsen/Germany for support of the construction of Advanced LIGO and construction and operation of the GEO600 detector. Additional support for Advanced LIGO was provided by the Australian Research Council. Virgo is funded, through the European Gravitational Observatory (EGO), by the French Centre National de Recherche Scientifique (CNRS), the Italian Istituto Nazionale di Fisica Nucleare (INFN) and the Dutch Nikhef, with contributions by institutions from Belgium, Germany, Greece, Hungary, Ireland, Japan, Monaco, Poland, Portugal, Spain. The construction and operation of KAGRA are funded by Ministry of Education, Culture, Sports, Science and Technology (MEXT), and Japan Society for the Promotion of Science (JSPS), National Research Foundation (NRF) and Ministry of Science and ICT (MSIT) in Korea, Academia Sinica (AS) and the Ministry of Science and Technology (MoST) in Taiwan.

\bibliography{Refs}

\onecolumngrid
\appendix

\section{Study of the FAP for the bivariate complex Gaussian}
\label{sec:anex:FAP_bcG}

\begin{figure*}[h!]
\centering
\includegraphics[width=0.49\textwidth]{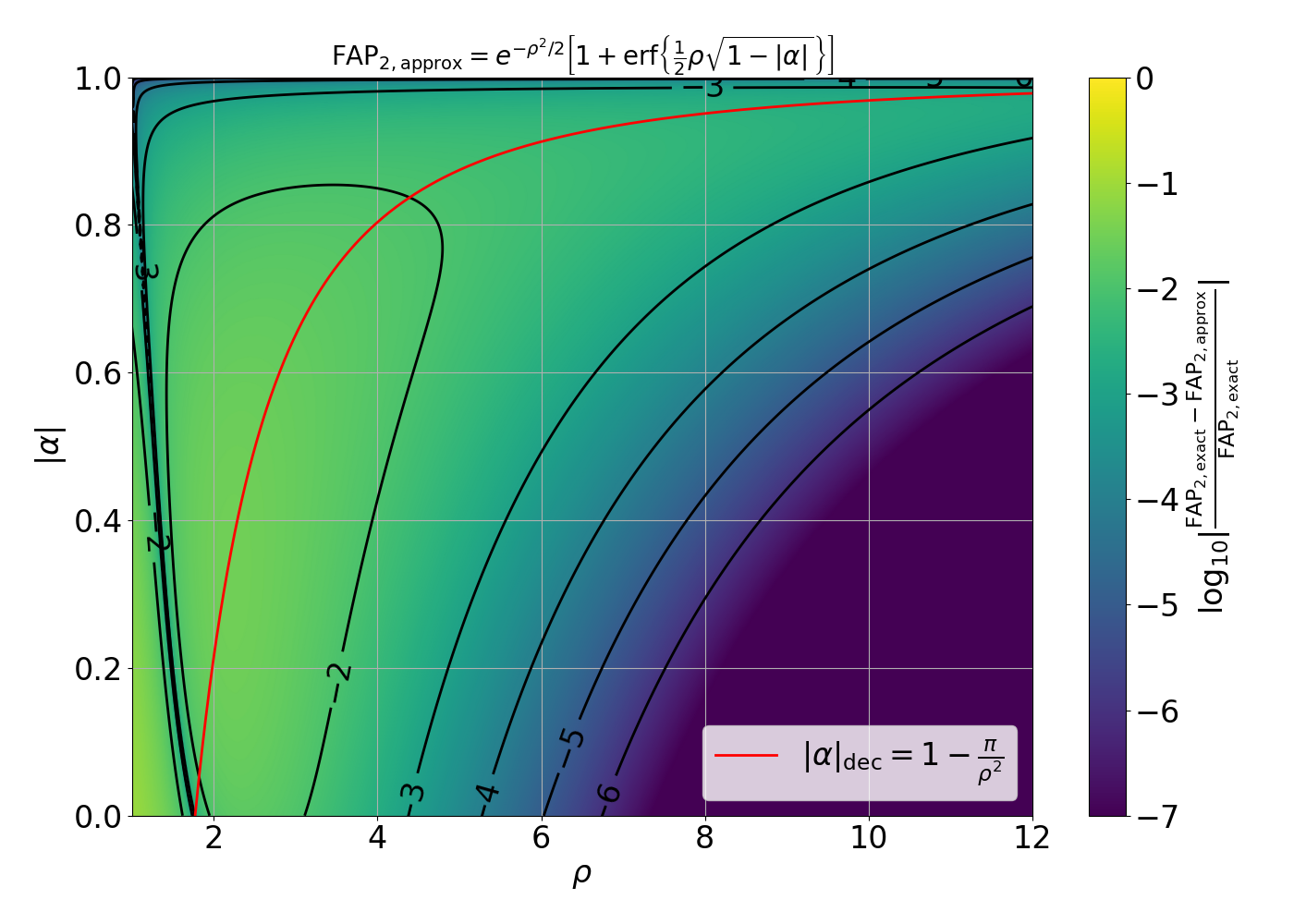}
\includegraphics[width=0.49\textwidth]{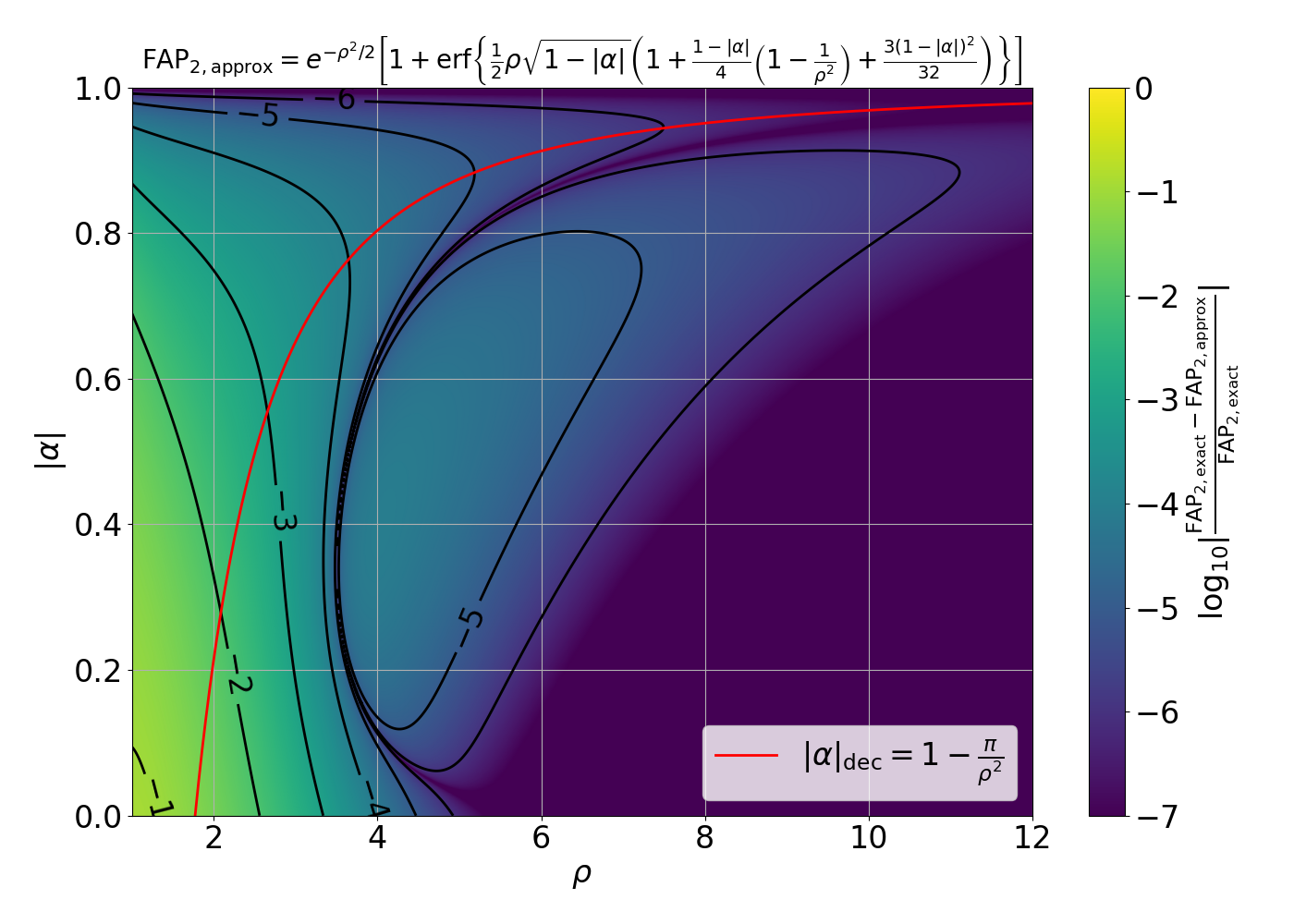}
\caption[Short title:group delay]{Base 10 logarithm of the relative error between the exact value of FAP$_2$, computed using Eq.~\eqref{eq:FAP2_7} and the approximations proposed in Eq.~\eqref{eq:FAP2_approx_LO} (left panel) and Eq.~\eqref{eq:FAP2_approx_NLO} (right panel), as a function of the correlation $|\alpha|$ and the SNR threshold $\rho$. We also show the with a red line the value of $|\alpha|$ at the decoupling time, to get an idea of the region where we are interested in having a good approximation as a function of $\rho$.}
\label{fig:FAP2_comparissons}
\end{figure*}

In this section we will study the FAP for the bivariate complex Gaussian (FAP$_2$) whose probability density function is given in Eq.~\eqref{eq:bivariate_cgaussian}. We will obtain Eq.~\eqref{eq:FAP2_7} to numerically compute FAP$_2$ in an efficient and well behaved manner. We also obtain a prescription to analytically approximate the FAP$_2$ to arbitrary order in $1-|\alpha|$ using Eq.~\eqref{eq:FAP2_7_approx_all_order}. With this expansion we obtain the leading order and second order approximations of Eq.~\eqref{eq:FAP2_approx_LO} and Eq.~\eqref{eq:FAP2_approx_NLO} respectively and shown in Fig.~\ref{fig:FAP2_comparissons}. 
As seen in Eq.~\eqref{eq:FAP2_0}, FAP$_2$ is given by the following expression:

\begin{align}
    \mathrm{FAP}_2 & = P(\rho_1 > \rho \, \cup \, \rho_2 > \rho) = 1 - P(\rho_1 < \rho \, \cap \, \rho_2 < \rho) \nonumber \\
    & = 1 - \frac{1}{(2\pi)^2 (1-|\alpha|^2)} \int_0^{2\pi} d \theta_1 \int_0^{2\pi} d \theta_2 \int_0^{\rho} \rho_1 d \rho_1 \int_0^{\rho} \rho_2 d \rho_2 \exp \left\{- \frac{\rho_1^2 + \rho_2^2 - 2 |\alpha| \rho_1 \rho_2 \cos (\theta_\alpha - \theta_1 + \theta_2 )}{2(1-|\alpha|^2)} \right\} \nonumber \\
    & = 1 - \frac{1}{2 \pi (1-|\alpha|^2)} \int_0^{\rho} d \rho_1 \int_0^{\rho} d \rho_2 \rho_1 \rho_2 \exp \left\{- \frac{\rho_1^2 + \rho_2^2}{2(1-|\alpha|^2)} \right\} \int_0^{2\pi} d \theta \exp \left\{\frac{|\alpha| \rho_1 \rho_2 }{1-|\alpha|^2} \cos{\theta} \right\} \nonumber \\
    & = 1 - \frac{1}{1-|\alpha|^2} \int_0^{\rho} d \rho_1 \int_0^{\rho} d \rho_2 \rho_1 \rho_2 \exp \left\{- \frac{\rho_1^2 + \rho_2^2}{2(1-|\alpha|^2)} \right\} I_0 \left\{ \frac{|\alpha|}{1-|\alpha|^2} \rho_1 \rho_2 \right\} \, ,
    \label{eq:FAP2_1}
\end{align}

\noindent where for notation simplicity we define $\alpha = \Gamma(\Delta t)$ and $I_n(z)$ is the modified Bessel function of the first kind \cite{Abramowitz_and_Stegun}:

\begin{equation}
    I_n(z) \equiv i^{-n} J_n(i x) = \frac{1}{\pi} \int_0^\pi d \theta \, e^{z \cos{\theta}} \cos(n\, \theta) = \sum_{k=0}^\infty \frac{\left(\frac{1}{2} z\right)^{2 k + n}}{k! (k+n)!} \quad (n \in \mathbb{Z}).
    \label{eq:I_n_z}
\end{equation}

The integral of Eq.~\eqref{eq:FAP2_1} can be further simplified by making the change of variables:

\begin{equation}
    \rho_i = \sqrt{2 (1-|\alpha|^2) u_i} \, \longrightarrow \,d \rho_i = \sqrt{\frac{1-|\alpha|^2}{2 u_i}} \, ,
    \label{eq:FAP2_var_change}
\end{equation}

\noindent which yields:

\begin{equation}
    \mathrm{FAP}_2 = 1 - (1-|\alpha|^2) \int_0^{x} d u_1 \int_0^{x} d u_2 I_0 ( 2 |\alpha| \sqrt{u_1 u_2} ) e^{-(u_1+u_2)} \, .    
    \label{eq:FAP2_2}
\end{equation}

\noindent where for notation simplicity we have defined:
\begin{equation}
    x \equiv \frac{\rho^2}{2(1-|\alpha|^2)} \, .
    \label{eq:x_FAP2_def}
\end{equation}

From Eq.~\eqref{eq:I_n_z} we have that the Taylor series of $I_0(z)$ around $z=0$ is given by:

\begin{equation}
    I_0(z) = \sum_{k=0}^\infty \frac{z^{2 k}}{2^{2k} (k!)^2} \, \rightarrow \, I_0(2 |\alpha| \sqrt{u_1 u_2}) = \sum_{k=0}^\infty  \frac{|\alpha|^{2 k} u_1^k u_2^k}{(k!)^2} \, .
    \label{eq:I_0_series}
\end{equation}

And substituting this expansion into Eq.~\eqref{eq:FAP2_2} we obtain:

\begin{equation}
    \mathrm{FAP}_2 = 1 - (1-|\alpha|^2) \sum_{k=0}^\infty |\alpha|^{2 k} \left[ \frac{1}{k!} \int_0^{x} u^k e^{-u} du \right]^2 \, .    
    \label{eq:FAP2_3}
\end{equation}

Since $k$ is a natural number, the integral appearing in Eq.~\eqref{eq:FAP2_3} is given by:

\begin{equation}
    \frac{1}{k!} \int_0^x u^k e^{-u} du = 
    1 - e^{-x} \sum_{n=0}^k \frac{x^n}{n!}
    \, ,
    \label{eq:integer_GammaLRI}
\end{equation}

Using this in Eq.~\eqref{eq:FAP2_3}, the FAP$_2$ will be given by:

\begin{align}
    \mathrm{FAP}_2 & = 1 - (1-|\alpha|^2) \sum_{k=0}^\infty |\alpha|^{2 k} \left[ 1 - e^{-x} \sum_{n=0}^k \frac{x^n}{n!} \right]^2 \nonumber \\
    & = 1 - (1-|\alpha|^2) \left[\sum_{k=0}^\infty |\alpha|^{2 k}  - 2 e^{-x} \sum_{k=0}^\infty \sum_{n=0}^k |\alpha|^{2 k} \frac{x^n}{n!} + e^{-2 x} \sum_{k=0}^\infty \sum_{n=0}^k \sum_{m=0}^k |\alpha|^{2 k} \frac{x^{n+m}}{n!m!}  \right] \, .    
    \label{eq:FAP2_4}
\end{align}

In the first sum of Eq.~\eqref{eq:FAP2_4} we recognize a simple geometric series. Taking into account that $|\alpha|^2<1$, it will converge to the following expression:

\begin{equation}
    \sum_{k=0}^\infty |\alpha|^{2 k} = \frac{1}{1-|\alpha|^2} \, .
    \label{eq:FAP2_4_S1}
\end{equation}

The second sum of Eq.~\eqref{eq:FAP2_4} can also be summed exactly by making some index manipulation:

\begin{equation}
    \sum_{k=0}^\infty |\alpha|^{2 k} \sum_{n=0}^k \frac{x^n}{n!} = \sum_{n=0}^\infty \frac{x^n}{n!} \sum_{k=n}^\infty |\alpha|^{2 k} = \sum_{n=0}^\infty \frac{(|\alpha|^2 x)^n}{n!} \sum_{k=0}^\infty |\alpha|^{2 k} = e^{|\alpha|^2 x} \frac{1}{1-|\alpha|^2} \, .
    \label{eq:FAP2_4_S2}
\end{equation}

Finally, the third sum of Eq.~\eqref{eq:FAP2_4} can not be summed exactly, but it can be significantly simplified by making similar index manipulations:

\begin{align}
    \sum_{k=0}^\infty |\alpha|^{2 k} \sum_{n=0}^k \sum_{m=0}^k \frac{x^{n+m}}{n!m!} & = \sum_{n=0}^\infty \sum_{m=0}^\infty \frac{x^{n+m}}{n!m!} \sum_{k=\mathrm{max}(n,m)}^\infty |\alpha|^{2 k}  = \sum_{k=0}^\infty |\alpha|^{2 k} \sum_{n=0}^\infty \sum_{m=0}^\infty |\alpha|^{2 \mathrm{max}(n,m)} \frac{x^{n+m}}{n!m!} \nonumber \\
    & = \frac{1}{1-|\alpha|^2} \sum_{n=0}^\infty \sum_{m=0}^\infty |\alpha|^{2 \mathrm{max}(n,m)} \frac{x^{n+m}}{n!m!} \, .
    \label{eq:FAP2_4_S3}
\end{align}

Substituting the results of the sums of Eqs.~\eqref{eq:FAP2_4_S1},~\eqref{eq:FAP2_4_S2},~\eqref{eq:FAP2_4_S3} into Eq.~\eqref{eq:FAP2_4}, we obtain the following result:

\begin{align}
    \mathrm{FAP}_2 = 2 e^{-(1-|\alpha|^2)x} - e^{-2x} \sum_{n=0}^\infty \sum_{m=0}^\infty |\alpha|^{2 \mathrm{max}(n,m)} \frac{x^{n+m}}{n!m!} \, .
    \label{eq:FAP2_5}
\end{align}

To further simplify this expression we can change indices in the sum of Eq.~\eqref{eq:FAP2_5}, using $l=n-m$ and $k=\frac{1}{2}(n+m)$:

\begin{equation}
    \sum_{n=0}^\infty \sum_{m=0}^\infty |\alpha|^{2 \mathrm{max}(n,m)} \frac{x^{n+m}}{n!m!} = \sum_{l=-\infty}^{\infty} \sum_{k=|l|/2}^{\infty} |\alpha|^{2 k + |l|} \frac{x^{2 k}}{\left(k + \frac{l}{2}\right)!\left(k - \frac{l}{2}\right)!} = S_0 + 2 \sum_{l=1}^{\infty} S_l \, , 
    \label{eq:FAP2_5_S}
\end{equation}

\noindent where we have used that $2\, \mathrm{max}(n,m) = n+m + |n-m| = 2k + |l|$ and we have defined:

\begin{equation}
    S_l = \sum_{k=l/2}^{\infty} |\alpha|^{2 k + l} \frac{x^{2 k}}{\left(k + \frac{l}{2}\right)!\left(k - \frac{l}{2}\right)!} = |\alpha|^l \underbrace{\sum_{k=0}^\infty \frac{(|\alpha| x)^{2 k + l}}{\left(k + l\right)!k!}}_{I_l(2 |\alpha| x)} = |\alpha|^l I_l(2 |\alpha| x) \, ,
    \label{eq:FAP2_5_S_Sl}
\end{equation}

\noindent where we have identified the Taylor series of the modified Bessel function of the first kind of order $l$ shown in Eq.~\eqref{eq:I_n_z}. Using Eq.~\eqref{eq:FAP2_5_S_Sl} and Eq.~\eqref{eq:FAP2_5_S} we have that the FAP$_2$ of Eq.~\eqref{eq:FAP2_5} will be given by:

\begin{align}
    \mathrm{FAP}_2 = 2 e^{-(1-|\alpha|^2)x} - e^{-2x} \left( I_0(2 |\alpha| x) + 2 \sum_{n=1}^\infty |\alpha|^n I_n (2 |\alpha| x) \right) \, .
    \label{eq:FAP2_6}
\end{align}

To compute the sum of modified bessel functions of the first kind, we can use their integral representation, shown in Eq.~\eqref{eq:I_n_z}:

\begin{align}
    I_0(z) + 2 \sum_{n=1}^\infty |\alpha|^n I_n (z) & = \frac{1}{\pi} \int_0^\pi d \theta \, e^{z \cos{\theta}} \left[ 1 + 2 \sum_{n=1}^\infty |\alpha|^n \cos(n\, \theta) \right] = \frac{1}{\pi} \int_0^\pi d \theta \, e^{z \cos{\theta}} \left[ 1 + \sum_{n=1}^\infty \left(|\alpha| e^{i\theta}\right)^n + \left(|\alpha| e^{-i\theta}\right)^n \right] \nonumber \\
    & = \frac{1}{\pi} \int_0^\pi d \theta \, e^{z \cos{\theta}} \left[ 1 +  \frac{|\alpha|e^{i\theta}}{1 - |\alpha|e^{i\theta}} + \frac{|\alpha|e^{-i\theta}}{1 - |\alpha|e^{-i\theta}} \right] = \frac{1}{\pi} \int_0^\pi d \theta \, e^{z \cos{\theta}} \frac{1-|\alpha|^2}{1 - 2|\alpha|\cos{\theta} + |\alpha|^2} \, .
    \label{eq:Bessel_sum_exact}
\end{align}

And we have transformed the infinite sum in a definite integral of a relatively simple function. The integral can be expressed in a more simple and convenient way if we do the variable change $\theta = 2 \arctan\left(\frac{1 - |\alpha|}{1 + |\alpha|} u\right)$:

\begin{align}
    I_0(z) + 2 \sum_{n=1}^\infty |\alpha|^n I_n (z) & = \frac{2}{\pi} e^z \int_0^\infty d u \frac{1}{1+u^2} \exp\left\{ - 2 z \frac{(1 - |\alpha|)^2 u^2}{(1 + |\alpha|)^2 + (1 - |\alpha|)^2 u^2}\right\} \, .
    \label{eq:Bessel_sum_exact_variable_change}
\end{align}

Substituting this expression for the sum into Eq.~\eqref{eq:FAP2_6} for the FAP$_2$ and using the fact that $z = 2 |\alpha| x$, where $x$ is defined in Eq.~\eqref{eq:x_FAP2_def}, we obtain:

\begin{align}
    \mathrm{FAP}_2 = 2 e^{-\rho^2/2} - \frac{2}{\pi} e^{-\rho^2/(1+|\alpha|)} \int_0^\infty d u \frac{1}{1+u^2} \exp\left\{- \frac{2 |\alpha|(1-|\alpha|) \rho^2}{(1 + |\alpha|)^3} \frac{u^2}{1 + (\frac{1 - |\alpha|}{1 + |\alpha|})^2 u^2}\right\} \, .
    \label{eq:FAP2_7}
\end{align}

The integral in this expression can not be analytically computed, but it can be numerically integrated as it is a well behaved one variable definite integral that does not suffer from divergences or accuracy problems due to large cancellations, as the previous integrals did. We can check that this formula has the correct limiting behavior if we realize that both when $|\alpha|=0$ and when $|\alpha|=1$, the argument of the exponential inside the integral of Eq.~\eqref{eq:FAP2_7} vanishes and the value of the integral is $\pi$/2. Therefore in the case in which $|\alpha|=0$, when there is no correlation, $\mathrm{FAP}_2(|\alpha|=0) = 2e^{-\rho^2/2} - e^{-\rho^2}  = 1 - (1-e^{-\rho^2/2})^2$ as is expected from two uncorrelated variables. In the opposite limit, when the correlation is maximal and $|\alpha| = 1$, FAP$_2$ coincides with the expected result in which the two variables behave as a single one, that is, $\mathrm{FAP}_2(|\alpha|=1) = e^{-\rho^2/2} = 1 - (1-e^{-\rho^2/2})^1$.

As seen in Sec.~\ref{sec:SNR_FAR} of the main text, we are interested in obtaining an approximation in the limit in which the correlation is large and thus $|\alpha| \to 1$. However, we will take into account that the SNR threshold $\rho$ can be large in such a way that $(1 - |\alpha) \rho^2$ can be of order $O(1)$. In this case, an upper bound approximation for the FAP$_2$ is obtained in the following way:

\begin{align}
    \mathrm{FAP}_2 & \approx 2 e^{-\rho^2/2} - \frac{2}{\pi} e^{-\rho^2/(1+|\alpha|)} \int_0^\infty d u \frac{1}{1+u^2} \exp\left\{- \frac{2 |\alpha|(1-|\alpha|) \rho^2}{(1 + |\alpha|)^3} u^2 \right\} \nonumber \\
    & = e^{-\rho^2/2} \left[ 2 - \exp\left\{-\frac{1}{2} \left(\frac{1-|\alpha|}{1+|\alpha|}\right)^3 \rho^2\right\} \mathrm{Erfc}\left\{ \rho \sqrt{\frac{2 |\alpha| (1-|\alpha|)}{(1+|\alpha|)^3}} \right\} \right] \nonumber \\
    & \approx e^{-\rho^2/2} \left[ 1 + \mathrm{Erf}\left\{ \frac{1}{2}\rho \sqrt{1 - |\alpha|} \right\} \right] \, ,
    \label{eq:FAP2_approx_LO}
\end{align}

\noindent where we have used that \cite{Abramowitz_and_Stegun}:

\begin{align}
    \frac{2}{\pi} \int_0^\infty  \frac{du}{1+u^2} e^{-\eta^2 u^2} = e^{\eta^2}\mathrm{Erfc}(\eta) \ \, ,
    \label{eq:erfc_integral}
\end{align}

\noindent and where $\mathrm{Erf}(z)$ and $\mathrm{Erfc}(z)$ are the error function and the complementary error function respectively. 
Eq.~\eqref{eq:FAP2_approx_LO} can be taken to be as the leading order term in an expansion in $1-|\alpha|$ of the FAP$_2$. To analyze higher order terms it will be convenient to introduce two new variables:

\begin{subequations}
\label{eq:eta_epsilon_def}
\begin{align}
    \eta & = \rho \sqrt{\frac{2 |\alpha| (1-|\alpha|)}{(1+|\alpha|)^3}} \, , \label{eq:eta_epsilon_def:eta} \\
    \epsilon & = \frac{1-|\alpha|}{1+|\alpha|} \, . \label{eq:eta_epsilon_def:epsilon}
\end{align}
\end{subequations}

In the regime we are interested, $\eta$ is of order $O(1)$, while $\epsilon \ll 1$. Using these variables we have:

\begin{align}
    \mathrm{FAP}_2 & = e^{-\rho^2/2} \left[2 - e^{- \frac{1-|\alpha|}{2 (1 + |\alpha|)} \rho^2} \frac{2}{\pi} \int_0^\infty  \frac{d u}{1+u^2}\exp\left\{-\frac{\eta^2 u^2}{1+\epsilon^2 u^2}\right\} \right] \nonumber \\
    & = e^{-\rho^2/2} \left[2 - e^{- \frac{1-|\alpha|}{2 (1 + |\alpha|)} \rho^2} \frac{2}{\pi} \int_0^\infty  \frac{d u}{1+u^2}e^{-\eta^2 u^2}\exp\left\{\frac{\epsilon^2 \eta^2 u^4}{1+\epsilon^2 u^2}\right\} \right] \nonumber \\
    & = e^{-\rho^2/2} \left[2 - e^{- \frac{1-|\alpha|}{2 (1 + |\alpha|)} \rho^2} \frac{2}{\pi} \int_0^\infty  \frac{d u}{1+u^2} e^{-\eta^2 u^2} \sum_{n=0}^\infty \frac{1}{n!} \left( \frac{\epsilon^2 \eta^2 u^4}{1+\epsilon^2 u^2} \right)^n \right] \, .
    \label{eq:FAP2_7_approx_all_order}
\end{align}

If we truncate the sum at $n$-th order, we obtain an upper bound approximation that is accurate to order $(\eta \epsilon)^{2 n}$ and that has correct limiting behavior when $\epsilon \to 0$, when $\epsilon = 1$, when $\eta = 0$ and when $\eta \to \infty$. Since we want only the first order correction, we can keep terms up to $n=1$ and integrate, obtaining: 
\begin{align}
    \mathrm{FAP}_2 & \approx  e^{-\rho^2/2} \left[2 - e^{- \frac{1-|\alpha|}{2 (1 + |\alpha|)} \rho^2} \frac{2}{\pi} \int_0^\infty  \frac{d u}{1+u^2} e^{-\eta^2 u^2} \left( 1 + \frac{\epsilon^2 \eta^2 u^4}{1+\epsilon^2 u^2} \right) \right] \nonumber \\
    & =  e^{-\rho^2/2} \left[2 - e^{- \frac{1-|\alpha|}{2 (1 + |\alpha|)} \rho^2} \left( \left( 1 + \frac{\epsilon^2 \eta^2}{1 - \epsilon^2}\right) e^{\eta^2} \mathrm{Erfc}(\eta) + \frac{\eta}{\sqrt{\pi}} - \frac{\eta^2}{\epsilon (1 - \epsilon^2)} e^{\eta^2/\epsilon^2} \mathrm{Erfc}\left(\frac{\eta}{\epsilon} \right)  \right) \right] \nonumber \\
    & \approx  e^{-\rho^2/2} \left[2 - e^{- \frac{1-|\alpha|}{2 (1 + |\alpha|)} \rho^2} \left( \left( 1 + \epsilon^2 \eta^2 \right) e^{\eta^2} \mathrm{Erfc}(\eta) - \epsilon^2 \frac{\eta}{\sqrt{\pi}} \left( 1 - \frac{1}{2 \eta^2} \right) \right) \right] \, .
    \label{eq:FAP2_7_approx_1}
\end{align}

We can express this result in terms of the correlation $|\alpha|$ and the SNR threshold $\rho$ substituting the expressions for $\eta$ and $\epsilon$ of Eq.~\eqref{eq:eta_epsilon_def}. To be consistent in the approximation, we keep the two first orders in $1-|\alpha|$, assuming that $(1 - |\alpha) \rho^2$ is of order $O(1)$. Doing this we obtain:
\begin{align}
    \mathrm{FAP}_2 & \approx  e^{-\rho^2/2} \left[1 + \mathrm{Erf}\left\{\frac{1}{2}\rho\sqrt{1-|\alpha|}\left(1 + \frac{1 - |\alpha|}{4} - \frac{(1-|\alpha|)^2}{32} \right) \right\} - \frac{(1-|\alpha|)^{3/2}}{4 \sqrt{\pi} \rho} e^{-\frac{1}{4}(1-|\alpha|)\rho^2} \left(1 -\frac{(1-|\alpha|)\rho^2}{2} \right) \right] \nonumber \\
    & \approx  e^{-\rho^2/2} \left[1 + \mathrm{Erf}\left\{\frac{1}{2}\rho\sqrt{1-|\alpha|}\left(1 + \frac{1 - |\alpha|}{4} \left(1 - \frac{1}{\rho^2} \right) + \frac{3 (1 - |\alpha|)^2}{32} \right) \right\} \right] \, .
    \label{eq:FAP2_approx_NLO}
\end{align}

\noindent where for simplicity of the final result, in the last step we have introduced all the corrections inside the argument of the error function in a way that is consistent with the order of the approximation. We check that ignoring the higher order corrections in $1-|\alpha|$, we recover the leading order expression of \eqref{eq:FAP2_approx_LO}.

In Fig.~\ref{fig:FAP2_comparissons} we show the relative error, between the exact FAP$_2$ computed using Eq.~\eqref{eq:FAP2_7} and the approximations of Eq.~\eqref{eq:FAP2_approx_LO} (left panel) and Eq.~\eqref{eq:FAP2_approx_NLO} (right panel), as a function of the correlation $|\alpha|$ and the SNR threshold $\rho$. We observe that the leading order approximation (left panel), already gives an accurate description of the FAP$_2$, having sub-percent accuracy for $\rho \gtrsim 5$ and reproducing the exact result as $|\alpha| \to 1$. On the right hand panel we can see the effect of introducing the higher order correction, we observe that the description is now much improved, reaching an accuracy better than 1 part in 10000 for $\rho \gtrsim 4$ and describing much better the limit $|\alpha| \to 1$. If we wanted to approximate the FAP$_2$ to higher precision, we could take into account more terms in the sum of Eq.~\eqref{eq:FAP2_7_approx_all_order} and analytically integrate them using Eq.~\eqref{eq:erfc_integral}.

\end{document}

%% file: Far_table.tex
\begin{ruledtabular}\begin{tabular}{l c c c c c c c c}
Event  & IFOs & Pipeline & Search & Search & $p_\text{astro}$ & PE & Gaussian  & Gaussian FAP$_\mathrm{event}$ \\[-2mm] 
 & &  & SNR &  $\log_{10}(\mathrm{FAR}\cdot\mathrm{yr})$ &  & SNR & $\log_{10}(\mathrm{FAR}\cdot\mathrm{yr})$ & ($T_\mathrm{obs} = 1 \mathrm{yr}$)  \\ \hline
GW191103\_012549 & HL & PyCBC-BBH & 9.3 & -0.34 & 0.94 & $8.9^{+0.3}_{-0.5}$ & $ -6.40^{+1.92}_{-1.08} $ & $(1.99 \pm 0.24) \times 10^{-4}$ \\[-1mm]
\makebox[0pt][l]{\fboxsep0pt\colorbox{lightgray}{\mystrut\hspace*{1.0\linewidth}}}\!\!
GW191105\_143521 & HLV & PyCBC-broad & 9.8 & -1.92 & $>0.99$ & $9.7^{+0.3}_{-0.5}$ & $ -9.44^{+2.08}_{-1.32} $ & $ (2.6 \pm 1.4) \times 10^{-6}$ \\[-1mm]%No Change
GW191109\_010717 & HL & MBTA & 15.2 & -3.74 & $>0.99$ & $17.3^{+0.5}_{-0.5}$ & $ -54.23^{+3.34}_{-4.04} $ &$(1.96 \pm 0.56) \times 10^{-50} $  \\[-1mm]
\makebox[0pt][l]{\fboxsep0pt\colorbox{lightgray}{\mystrut\hspace*{1.0\linewidth}}}\!\!
GW191113\_071753 & HLV & MBTA & 9.2 & 1.41 & 0.68 & $ 7.8^{+0.6}_{-1.1} $ & $ -2.40^{+3.35}_{-2.14} $ & $  0.15724 \pm 0.00085$ \\[-1mm]
GW191126\_115259 & HL & PyCBC-BBH & 8.5 & 0.51 & 0.70 & $8.3^{+0.2}_{-0.5}$ &$ -4.27^{+1.68}_{-0.76} $ & $ (4.05 \pm 0.12) \times 10^{-3} $ \\[-1mm]
\makebox[0pt][l]{\fboxsep0pt\colorbox{lightgray}{\mystrut\hspace*{1.0\linewidth}}}\!\!
GW191127\_050227 & HLV & PyCBC-BBH & 8.7 & 0.61 & 0.74 & $9.1^{+0.5}_{-0.6}$ & $ -7.77^{+2.29}_{-2.23} $ & $ (2.35 \pm 0.54) \times 10^{-5} $ \\[-1mm]
GW191129\_134029 & HL & GstLAL & 13.3 & $<-5$ & $>0.99$ & $13.2^{+0.2}_{-0.3}$ & $ -26.58^{+1.60}_{-1.21} $ & $  (1.90 \pm 0.36) \times 10^{-25} $ \\[-1mm]
\makebox[0pt][l]{\fboxsep0pt\colorbox{lightgray}{\mystrut\hspace*{1.0\linewidth}}}\!\!
GW191204\_110529 & HL & PyCBC-BBH & 8.9 & 0.52 & 0.74 & $8.8^{+0.4}_{-0.6}$ & $ -6.15^{+2.28}_{-1.62} $ & $ (4.68 \pm 0.36) \times 10^{-4} $ \\[-1mm]
GW191204\_171526 & HL & PyCBC-broad & 17.1 & $<-5$ & $>0.99$ & $17.5^{+0.2}_{-0.2}$ & $ -55.15^{+1.80}_{-1.40} $ & $(1.7 \pm 1.4) \times 10^{-52} $ \\[-1mm]
\makebox[0pt][l]{\fboxsep0pt\colorbox{lightgray}{\mystrut\hspace*{1.0\linewidth}}}\!\!
GW191215\_223052 & HLV & GstLAL & 10.9 & $<-5$ & $>0.99$ & $11.2^{+0.3}_{-0.4}$ &  $ -16.38^{+2.00}_{-1.59} $ & $ (9.5 \pm 2.7) \times 10^{-15} $\\[-1mm]
GW191216\_213338 & HV & GstLAL & 18.6 & $<-5$  & $>0.99$ & $18.6^{+0.2}_{-0.2}$ & $ -63.74^{+1.81}_{-1.47} $ & $ (8.1 \pm 2.3) \times 10^{-62} $ \\[-1mm]
\makebox[0pt][l]{\fboxsep0pt\colorbox{lightgray}{\mystrut\hspace*{1.0\linewidth}}}\!\!
GW191219\_163120 & HLV & PyCBC-broad & 8.9 & 0.60 & 0.82 & $9.1^{+0.5}_{-0.8}$ & $ -7.61^{+3.00}_{-2.07} $  & $ (2.29 \pm 0.33) \times 10^{-3} $ \\[-1mm]
GW191222\_033537 & HL & GstLAL & 12 & $<-5$ & $>0.99$ & $12.5^{+0.2}_{-0.3}$ & $ -23.29^{+1.53}_{-1.12} $ & $ (2.2 \pm 2.0) \times 10^{-21} $ \\[-1mm]
\makebox[0pt][l]{\fboxsep0pt\colorbox{lightgray}{\mystrut\hspace*{1.0\linewidth}}}\!\!
GW191230\_180458 & HLV & PyCBC-BBH & 9.9 & -0.38 & 0.96 & $ 10.5^{+0.2}_{-0.4} $ & $ -13.48^{+1.74}_{-1.09} $& $(3.6 \pm 3.5) \times 10^{-10}$ \\[-1mm]
GW200112\_155838 & LV & GstLAL & 17.6 &  $<-5$  & $>0.99$ & $19.8^{+0.1}_{-0.2}$ & $ -74.28^{+1.79}_{-1.17} $ & $ (1.82 \pm 0.79) \times 10^{-72} $ \\[-1mm]
\makebox[0pt][l]{\fboxsep0pt\colorbox{lightgray}{\mystrut\hspace*{1.0\linewidth}}}\!\!
GW200115\_042309 & HLV & GstLAL & 11.5 & $<-5$ & $>0.99$ & $11.3^{+0.3}_{-0.5}$ & $ -16.69^{+2.43}_{-1.51} $ & $(8.1 \pm 5.1) \times 10^{-14} $ \\[-1mm]
GW200128\_022011 & HL & PyCBC-BBH & 9.9 & -2.37 & $>0.99$ & $10.7^{+0.3}_{-0.4}$ & $ -14.16^{+1.66}_{-1.38} $ & $ (4.92 \pm 0.78) \times 10^{-13} $\\[-1mm]
\makebox[0pt][l]{\fboxsep0pt\colorbox{lightgray}{\mystrut\hspace*{1.0\linewidth}}}\!\!
GW200129\_065458 & HLV & GstLAL & 26.5 & $<-5$ & $>0.99$ & $26.8^{+0.2}_{-0.2}$  & $ -144.95^{+2.39}_{-2.21} $ &  $  (5.94 \pm 0.96) \times 10^{-143} $ \\[-1mm]
GW200202\_154313 & HLV & GstLAL & 11.3 & $<-5$ & $>0.99$ & $10.9^{+0.2}_{-0.4}$ & $ -14.63^{+1.76}_{-1.05} $ & $  (3.9 \pm 2.2) \times 10^{-11} $ \\[-1mm]
\makebox[0pt][l]{\fboxsep0pt\colorbox{lightgray}{\mystrut\hspace*{1.0\linewidth}}}\!\!
GW200208\_130117 & HLV & PyCBC-BBH & 10.8 & -3.51 & $>0.99$ & $10.9^{+0.2}_{-0.4}$& $ -15.04^{+1.96}_{-1.13} $ & $  (3.4 \pm 2.0) \times 10^{-11} $\\[-1mm]
GW200208\_222617 & HLV & PyCBC-BBH & 7.9 & 0.68 & 0.70 & $7.4^{+1.1}_{-2.0}$ & $ -1.41^{+5.28}_{-3.95} $  & $ 0.31395 \pm 0.00090 $ \\[-1mm]
\makebox[0pt][l]{\fboxsep0pt\colorbox{lightgray}{\mystrut\hspace*{1.0\linewidth}}}\!\!
GW200209\_085452 & HLV & MBTA & 9.7 & 1.08 & 0.97 & $9.6^{+0.3}_{-0.5}$ & $ -9.67^{+1.99}_{-1.38} $ & $  (2.3 \pm 1.9) \times 10^{-6} $  \\[-1mm]
GW200210\_092254 & HLV & PyCBC-BBH & 8.9 & 0.89 & 0.54 & $8.4^{+0.5}_{-0.7}$ & $ -4.66^{+2.50}_{-1.84} $ & $  (1.169 \pm 0.025) \times 10^{-2} $  \\[-1mm]
\makebox[0pt][l]{\fboxsep0pt\colorbox{lightgray}{\mystrut\hspace*{1.0\linewidth}}}\!\!
GW200216\_220804 & HLV & GstLAL & 9.4 & -0.45 & 0.77 & $ 8.2^{+0.3}_{-0.5} $ & $ -4.24^{+1.72}_{-1.05} $ & $  (2.948 \pm 0.095) \times 10^{-3} $  \\[-1mm]
GW200219\_094415 & HLV & GstLAL & 10.7 & -3.00 & $>0.99$ & $10.7^{+0.3}_{-0.4}$ & $ -14.45^{+1.98}_{-1.33} $  & $ (1.4 \pm 1.2) \times 10^{-11}$ \\[-1mm]
\makebox[0pt][l]{\fboxsep0pt\colorbox{lightgray}{\mystrut\hspace*{1.0\linewidth}}}\!\!
GW200220\_061928 & HLV & PyCBC-BBH & 7.5 & 0.83 & 0.62 & $7.3^{+0.4}_{-0.7}$ & $ -1.66^{+1.94}_{-1.13} $ & $  0.13003 \pm 0.00070$  \\[-1mm]
GW200220\_124850 & HL & MBTA & 8.2 & -2.74 & 0.83 & $8.5^{+0.3}_{-0.5}$ & $ -5.30^{+1.74}_{-1.02} $ & $ (5.41 \pm 0.47) \times 10^{-4}$  \\[-1mm]
\makebox[0pt][l]{\fboxsep0pt\colorbox{lightgray}{\mystrut\hspace*{1.0\linewidth}}}\!\!
GW200224\_222234 & HLV & MBTA & 19.0 & $<-5$ & $>0.99$ & $20.0^{+0.2}_{-0.2}$ & $ -75.77^{+1.84}_{-1.41} $ & $ (7.9 \pm 4.3) \times 10^{-74} $  \\[-1mm]
GW200225\_060421 & HL & PyCBC-broad & 12.3 & $<-5$ & $>0.99$ & $12.5^{+0.3}_{-0.4}$ & $ -23.14^{+1.88}_{-1.57} $ & $ (1.44 \pm 0.35) \times 10^{-21} $ \\[-1mm]
\makebox[0pt][l]{\fboxsep0pt\colorbox{lightgray}{\mystrut\hspace*{1.0\linewidth}}}\!\!
GW200302\_015811 & HV & GstLAL & 10.6 & -0.96 & 0.91 & $10.8^{+0.3}_{-0.4}$ &  $ -14.76^{+1.94}_{-1.64} $  & $  (3.11 \pm 0.76) \times 10^{-13} $ \\[-1mm]
GW200306\_093714 & HL & MBTA & 8.5 & 2.61 & 0.81 & $ 7.8^{+0.3}_{-0.6} $ & $ -2.46^{+1.95}_{-1.19} $  & $  (4.933 \pm 0.047) \times 10^{-2} $  \\[-1mm]
\makebox[0pt][l]{\fboxsep0pt\colorbox{lightgray}{\mystrut\hspace*{1.0\linewidth}}}\!\!
GW200308\_173609 & HLV & PyCBC-BBH & 8.0 & 0.38 & 0.86 & $ 3.8^{+3.1}_{-2.5} $ &  $ 6.55^{+1.91}_{-6.17} $ & $  0.96500 \pm 0.00045$  \\[-1mm]
\textit{GW200308\_173609}$^*$ & - & - & - & - & - & $ 7.09^{+0.47}_{-0.50} $ &  $ -0.90^{+3.34}_{-3.31} $ & $  0.4366 \pm 0.0040$  \\[-1mm]
\makebox[0pt][l]{\fboxsep0pt\colorbox{lightgray}{\mystrut\hspace*{1.0\linewidth}}}\!\!
GW200311\_115853 & HLV & GstLAL & 17.7 & $<-5$ & $>0.99$ & $ 17.9^{+0.1}_{-0.2} $ & $ -58.41^{+1.69}_{-1.12} $ & $  (6.7 \pm 1.8) \times 10^{-57} $  \\[-1mm]
GW200316\_215756 & HLV & GstLAL & 10.1 & $<-5$ &  $>0.99$ & $10.3^{+0.4}_{-0.7}$ &   $ -12.24^{+2.82}_{-1.81} $ & $  (2.5 \pm 2.0) \times 10^{-8} $  \\[-1mm]
\makebox[0pt][l]{\fboxsep0pt\colorbox{lightgray}{\mystrut\hspace*{1.0\linewidth}}}\!\!
GW200322\_091133 & HLV & MBTA & 9.0 & 2.65 & 0.62 & $ 2.5^{+3.4}_{-1.7} $ & $ 8.00^{+0.75}_{-5.35} $ &  $  0.99327 \pm 0.00021$ \\[-1mm]
\textit{GW200322\_091133}$^*$ & - & - & - & - & - & $ 5.3^{+1.4}_{-0.9} $ & $ 9.15^{+4.14}_{-7.37} $ &  $  0.96870 \pm 0.00096$ \\
\end{tabular}\end{ruledtabular}

% IFOs_search, Most significant pipeline, SNR_search, FAR_search, pastro, SNR_PE, FAR_Gaussiano 